\documentclass[preprints,review,accept,moreauthors,pdftex]{Definitions/mdpi} 

\firstpage{1} 
\pubvolume{xx}
\issuenum{xx}
\articlenumber{xx}
\pubyear{2021}
\copyrightyear{2021}
\history{Received: 16th Feb. 2021; Accepted: 4th March 2021; Published: 10th March 2021}

\pdfoutput=1

\usepackage{subcaption}
\usepackage{textcomp}

\Title{Review of Liquid Argon Detector Technologies in the Neutrino Sector}

\Author{Krishanu Majumdar * and Konstantinos Mavrokoridis *}

\AuthorNames{Krishanu Majumdar and Konstantinos Mavrokoridis}

\address{University of Liverpool, Department of Physics, Oliver Lodge Bld, Oxford Street, Liverpool, L69 7ZE, UK}

\corres{Correspondence: majumdar@liverpool.ac.uk (K.Mj.); k.mavrokoridis@liverpool.ac.uk (K.Mv.)}

\abstract{Liquid Argon (LAr) is one of the most widely used scintillators in particle detection, due to its low cost, high availability and excellent scintillation properties. A large number of experiments in the neutrino sector are based around using LAr in one or more Time Projection Chambers (TPCs), leading to high resolution three-dimensional particle reconstruction. In this paper, we review and summarise a number of these LArTPC experiments, and briefly describe the specific technologies that they currently employ.
This includes single phase LAr experiments (ICARUS~T600, MicroBooNE, SBND, LArIAT, DUNE-SP, ProtoDUNE-SP, ArgonCube and Vertical Drift) and dual phase LAr experiments (DUNE-DP, WA105, ProtoDUNE-DP and ARIADNE).
We also discuss some new avenues of research in the field of LArTPC readout, which show potential for wide-scale use in the near future.}


\begin{document}

{\hypersetup{linktocpage} \tableofcontents}
\noindent \hrulefill
\vspace{5mm}

\section{Introduction}

It has been ninety years since Wolfgang Pauli first proposed the existence of particles which he described as ``\emph{electrically neutral ... that have spin 1/2 and obey the exclusion principle and that further differ from light quanta in that they do not travel with the velocity of light}'' \cite{PauliNeutrinos}. We now of course know these as neutrinos, but despite nearly seventy years passing since the direct observation of the electron (anti)neutrino by Cowan and Reines \cite{CowanReines1953}, there is still much to be discovered in the neutrino sector of Particle Physics.

A common detection technique in neutrino experiments (as well as other sectors of Particle Physics) is the use of scintillators - materials that emit photons (``luminesce'') when exposed to ionising radiation \cite{GeneralScintillation1}. Scintillators come in a variety of forms: gases, liquids and solids, and depending on the specific material, scintillation light can be emitted at infrared, visible, UV and even x-ray wavelengths. 

One of the most important scintillators for neutrino detectors is liquid argon (LAr). Argon is the most abundant noble gas on Earth, constituting 0.934\% of the atmosphere's volume \cite{ArgonBritannica}, and is commercially extracted, liquefied and purified at large scales worldwide, making it relatively cheap and plentiful. It scintillates in the vacuum ultraviolet (``VUV'') wavelength range, with a peak emission at approximately 128~nm \cite{LArScintillation1, LArScintillation2}, and in its liquid phase has a relatively high scintillation yield (that is, the amount of light emitted per unit of energy deposited by the incident ionising radiation) of the order of 40,000 photons per MeV \cite{GeneralScintillation1, GeneralScintillation2}. LAr has been used as a scintillator in Particle Physics experiments since the late 1970s, following Carlo Rubbia's proposal for a LAr Time Projection Chamber (TPC) \cite{RubbiaLArTPC}. The general TPC concept had been realised a few years earlier \cite{NygrenGeneralTPC} as a solution to the simultaneous requirements of high precision position measurement and complete and unambiguous three-dimensional position reconstruction, but was originally envisioned as using a noble gas medium. However, given that particle interaction cross-sections scale with the density of the interacting material, the liquid phase of argon is a vastly more effective scintillator than its gaseous phase, the density of which is approximately one thousandth that of LAr \cite{LArProperties}.

LArTPCs continue to be at the forefront of experiments in the neutrino sector, growing in size, sophistication and capability in the past fifty years. There are now two main designs for such detectors: single-phase (which utilises only the liquid phase of argon) and dual-phase (which contains both liquid and gaseous forms). This paper will review the detection and design principles, primary hardware and status of a number of currently and recently operating single-phase (Section~\ref{sec:singlePhase}) and dual-phase (Section~\ref{sec:dualPhase}) LArTPCs, as well as a selection of developing technologies that could potentially be used in future experiments.

\section{Single-Phase Experiments}
\label{sec:singlePhase}
\begin{figure}[h]
\centering
\includegraphics[width=0.85\textwidth]{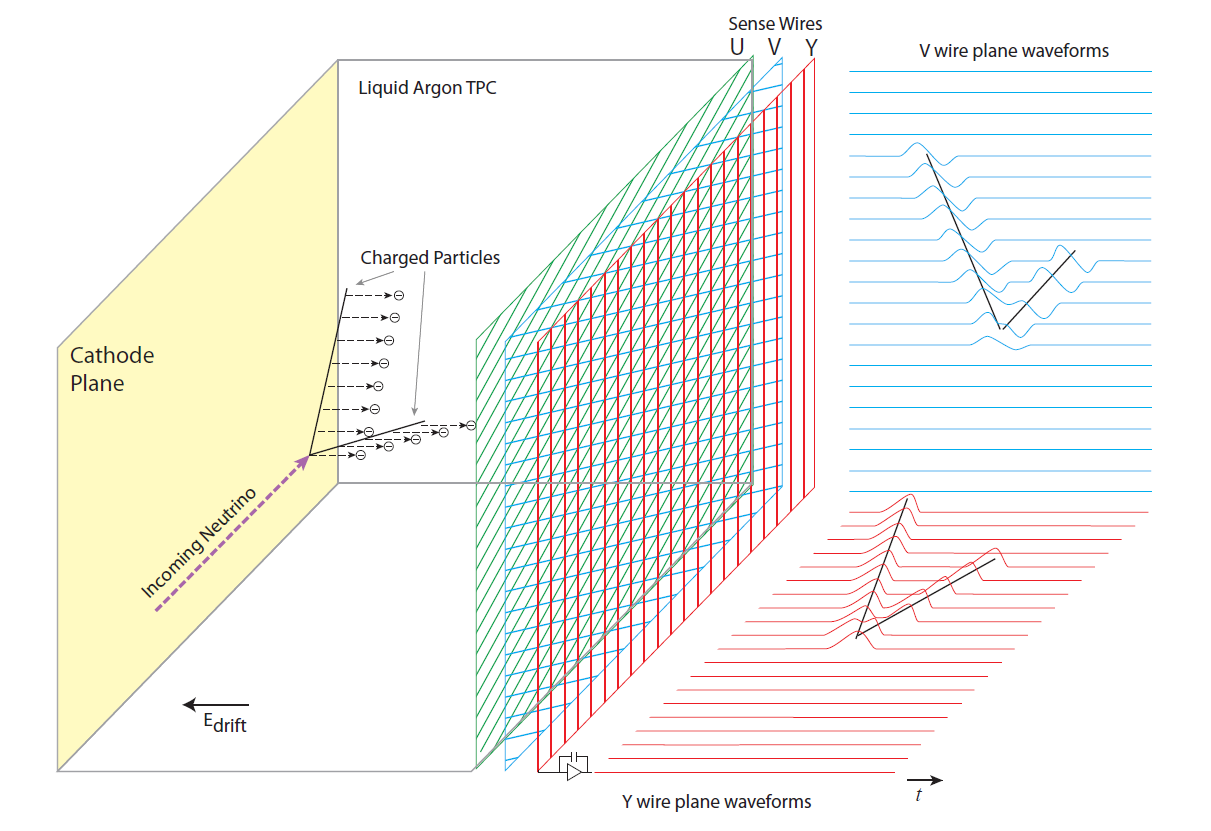}
\caption{The general detection principle of single-phase LAr detectors. This schematic depicts the assembly in the MicroBooNE detector, which uses a wire-based anode plane assembly for charge readout, but the same principle is broadly applicable to any single-phase design. A description of the operation is given in the main text. Taken from \cite{MicroBooNE_TDR}.}
\label{fig:MicroBooNE_DetectionPrinciple}
\end{figure}

All single-phase LArTPC detectors are broadly based on the same design as the original concept, and therefore share the same detection and operating principles, as described below and depicted in Figure~\ref{fig:MicroBooNE_DetectionPrinciple}. When a charged particle traverses the detector's active LAr volume, it both excites and ionises the argon. Such particles can enter the detector directly - for example, cosmic muons or charged particles from a beamline - or be produced by interactions within the LAr. An example of the latter is the elastic scattering of a neutrino from an atomic electron: on its own, the neutrino does not produce any ionisation, but the scattered electron does, therefore allowing the presence of the neutrino to be inferred. Excited Ar atoms de-excite on an extremely fast timescale, leading to the emission of ``prompt'' scintillation light, and the ionisation of the Ar atoms results in the production of free electrons along the particle's path. These electrons are drifted in a uniform electric field (the ``drift field'') towards one or more anode planes.

In detectors that utilise wire-based anode planes (which is the most common scheme used for charge readout), each plane is biased at a certain voltage that dictates its response to the electrons. One that is biased with a negative or zero voltage acts as an ``induction'' plane: bipolar voltage signals are induced on any wires that the electrons pass near to, with the amplitude of the signal being proportional to the distance between wire and electron, but the electrons will not be captured on the wires. However, if the plane is biased at a positive voltage, the electrons will be attracted to, and therefore collected on, the closest wire - thereby producing unipolar current signals on this so-called ``collection'' plane. When using multiple wire planes, the wires on each one are arranged in parallel, but orientated at different relative angles between planes, resulting in the ``crossing points'' between wires on different planes being unique and unambiguous - that is, no wire on a particular plane crosses over a wire on a different plane more than once \cite{WireReconstruction}. This means that the 2D position at which an electron is incident on the planes, and therefore the 2D projection of the ionisation point, can be localised to a single crossing point that only depends on the specific wires that signals are produced on. The third dimension of the ionisation point can be calculated using the time difference between the start of the event (determined via the detection of the aforementioned prompt scintillation light) and the signal production on the wires. When performed over all wire signals produced during a single event, a series of 3D ionisation points can therefore be determined, corresponding to the original particle track \cite{WireReconstruction}. Additionally, the total charge accumulated on the collection plane is directly related to the number of electrons produced by the original particle's ionisation, and this in turn is related to its energy.

Although wire-based anode planes are the most established type of charge readout in single-phase LArTPCs, alternate methods have been developed with great success, as described below.

\subsection{ICARUS T600}
Originally proposed in 1985 \cite{ICARUS_OrigProposal}, the ICARUS (``\textbf{I}maging \textbf{C}osmic \textbf{A}nd \textbf{R}are \textbf{U}nderground \textbf{S}ignals'') program has pioneered many of the techniques for single- and dual-phase LArTPC detector design, assembly and operation that are used by all other experiments operating today. Supported by a long-running research and development framework that included fully realised and operational 3 \cite{ICARUS_3tonDescr, ICARUS_3tonResults} and 14~tonne \cite{ICARUS_14ton} prototypes, the culmination of the program - the ICARUS T600 single-phase detector - was successfully assembled and operated at the Gran Sasso National Laboratories between 2004 and 2013 \cite{ICARUS_LNGSOperation, ICARUS_Papers}. Following this, the detector was refurbished and moved to Fermilab in 2017 \cite{ICARUS_FNALRefurb}, to take its place as the far detector of the Short Baseline Neutrino (SBN) Program \cite{SBN_Review}, with operation expected to begin in 2021.

\begin{figure}[h]
\centering
\includegraphics[width=0.82\textwidth]{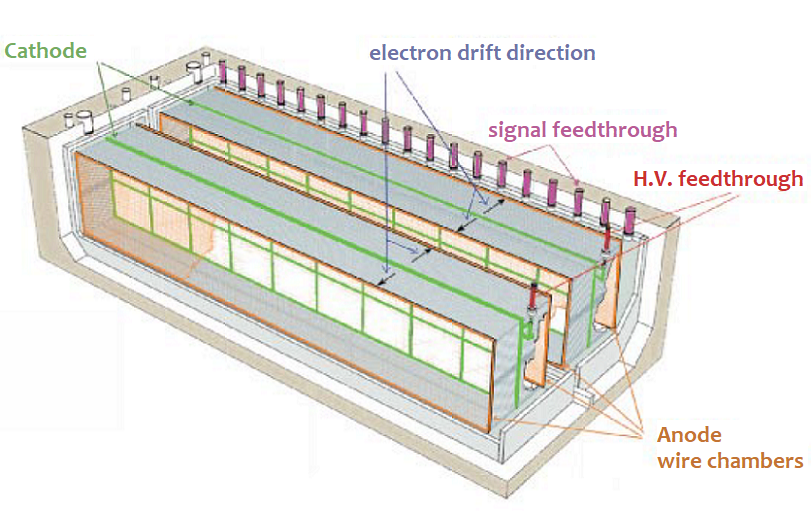}
\caption{A schematic depiction of the ICARUS T600 detector, with major components labelled. Taken from \cite{ICARUS_Image}.}
\label{fig:ICARUS_Detector}
\end{figure}

\begin{figure}[ht]
\centering
\includegraphics[width=0.82\textwidth]{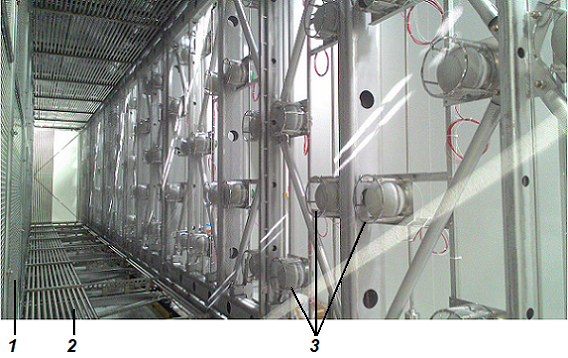}
\caption{A view inside one of the four ICARUS T600 drift volumes, showing 1) the cathode, 2) the field-shaping electrodes, and 3) the photomultiplier tube (``PMT'') array for detection of scintillation light. The three wire planes which make up the anode plane assembly are positioned in front of the PMTs, but the individual wires are too thin to be seen on this image. Adapted from \cite{ICARUS_PMTs}.}
\label{fig:ICARUS_TPC}
\end{figure}

The layout of ICARUS T600 is shown in Figure~\ref{fig:ICARUS_Detector}. The detector consists of two identical LAr modules, each with dimensions of 17.95 $\times$ 3 $\times$ 3.16~m (giving a combined active LAr mass of 476~tonnes) \cite{ICARUS_TDR}, enclosed in a single cryostat. Each module is split into two drift volumes, arranged ``back to back'' and sharing a common cathode orientated vertically and running lengthwise along the module. A view inside one of the drift volumes is shown in Figure~\ref{fig:ICARUS_TPC}. The entire cryostat is surrounded by cosmic ray tagger panels that are used to correlate incident cosmic muons with their corresponding signals in the active LAr \cite{ICARUS_FNALRefurb}.

\begin{figure}[h!]
\vspace{5mm}
\centering
\includegraphics[width=0.48\textwidth]{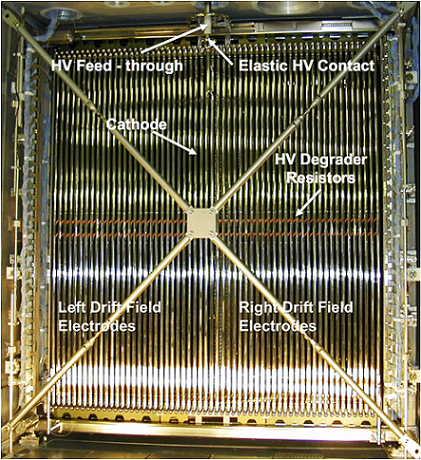}
\caption{An end-on view of one of the ICARUS T600 modules, showing the various high voltage components servicing two of the drift volumes. Taken from \cite{ICARUS_Components}.}
\label{fig:ICARUS_HVComponents}
\end{figure}

Each cathode is made of nine punched stainless steel panels, and nominally biased to -75~kV via a bespoke high voltage (``HV'') feedthrough \cite{ICARUS_TDR}. Between the cathode and anode planes, the drift field is created through the use of 29 ``race track'' field-shaping electrodes working in tandem with an attached resistor chain. The distance between the cathode and innermost anode plane - i.e. the drift length - is 1.5~m, and during normal operation the drift field is set at 500~V/cm, corresponding to a maximum drift time in ultra-pure LAr of 950~$\mu$s \cite{ICARUS_TDR}. A view of the HV components in situ is shown in Figure~\ref{fig:ICARUS_HVComponents}.

\begin{figure}[htp]
\centering
\includegraphics[width=0.48\textwidth]{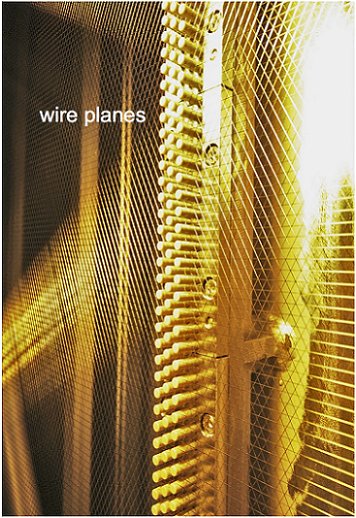}
\caption{A close-up view of the three wire planes used by one of the ICARUS T600 drift volumes. Taken from \cite{ICARUS_Components}.}
\label{fig:ICARUS_Wires}
\end{figure}

Each of the ICARUS T600 drift volumes utilises three wire-based anode planes for charge readout. (For consistency with other detectors discussed in this paper, the entire three-plane anode structure is referred to as the ``anode plane assembly'' (APA).)  Across all four APAs used in ICARUS T600, there are a total of 53,248 wires \cite{ICARUS_TDR}, with 3~mm spacing between wires and 3~mm spacing between the planes in a single APA. The wires are orientated at 0\textdegree{} (horizontal) on the innermost plane  (i.e. the one closest to the cathode), +60\textdegree{} for the middle plane, and -60\textdegree{} for the outermost plane. The two inner planes are nominally biased at -220 and 0~V respectively, making them induction planes, and the outermost plane operates in collection mode, being biased at +280~V. Figure~\ref{fig:ICARUS_Wires} shows a close-up view of the wire planes installed in one of the drift volumes.

\begin{figure}[hbp]
\vspace{5mm}
\centering
\includegraphics[width=0.60\textwidth]{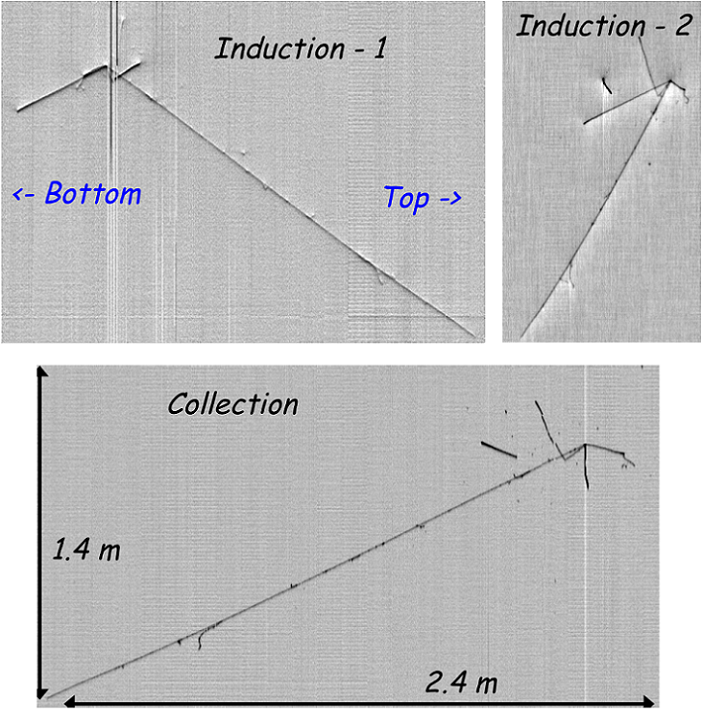}
\caption{Two-dimensional views of a single upward-going atmospheric muon neutrino event as seen on the 0\textdegree{} and +60\textdegree{} induction wire planes (top left and top right respectively) and the -60\textdegree{} collection plane (bottom) of one of the ICARUS T600 APAs. Adapted from \cite{ICARUS_ReadoutElecs}.}
\label{fig:ICARUS_Event2DViews}
\end{figure}

Wire signal acquisition and recording are performed using a system of dedicated ADCs and motherboards housed in custom-built electronics crates. Rather than being located off-detector (which would introduce additional cabling and other complexities), these crates are directly mounted to the external sides of the flanges that cap the detector's 96 signal feedthrough chimneys (previously depicted in Figure~\ref{fig:ICARUS_Detector}). A custom-designed PCB built into each flange acts as the electronic interface between the external and internal sides, with front-end electronics - such as amplifiers and multiplexing capabilities (that is, the collation of multiple signals for transmission over a single line) - being housed on dedicated motherboards mounted on the internal side of each PCB. A total of nine ``decoupling and bias boards'' (DBBs) connect to each front-end motherboard, and these both provide the bias to and receive signal readout from the wires, with each DBB servicing 64 wires \cite{ICARUS_ReadoutElecs}. As noted previously, each plane produces a single 2D view of an event specific to that plane's wire orientation, and by combining these views, 3D position reconstruction is achieved. An example of the wire plane views from a single upward-going atmospheric muon neutrino event is shown in Figure~\ref{fig:ICARUS_Event2DViews}.

Behind each drift volume's APA, in the non-active LAr region, are situated an array of photomultiplier tubes (``PMTs'') which are used to detect the prompt scintillation light. This light is produced effectively instantaneously upon the particle's entry into the active LAr volume, and so its detection time denotes the ``start'' of the event. Each drift volume is equipped with 90 8-inch diameter Hamamatsu R5912-MOD PMTs \cite{ICARUS_PMTs}, arranged in a grid as previously shown in Figure~\ref{fig:ICARUS_TPC}. Each PMT's outer surface is coated with Tetra-Phenyl Butadiene (TPB) - a wavelength shifter that converts the VUV scintillation light into photons in the visible wavelength range \cite{TPB_Emission}, where the PMTs are most sensitive. To avoid the possibility of the PMTs inducing signals on the wire planes, each PMT is housed in a grounded stainless steel cage \cite{ICARUS_PMTs}.

\subsection{MicroBooNE}
In a similar vein to ICARUS T600, the MicroBooNE (``\textbf{Micro} \textbf{Boo}ster \textbf{N}eutrino \textbf{E}xperiment'') detector is another previously independent single-phase LArTPC that has now been re-purposed as part of the SBN Program. Originally designed and built primarily to investigate the anomalous excess of electron-like events observed in a low energy neutrino beam by its predecessor MiniBooNE \cite{MiniBooNE_Results}, MicroBooNE has operated on the Booster Neutrino Beamline at Fermilab since 2015 \cite{MicroBooNE_TDR} - independently until 2018, and since then falling under the umbrella of the SBN Program, within which it is positioned as the intermediate detector \cite{SBN_Review}.

\begin{figure}[h]
\vspace{2mm}
\centering
\includegraphics[width=0.60\textwidth]{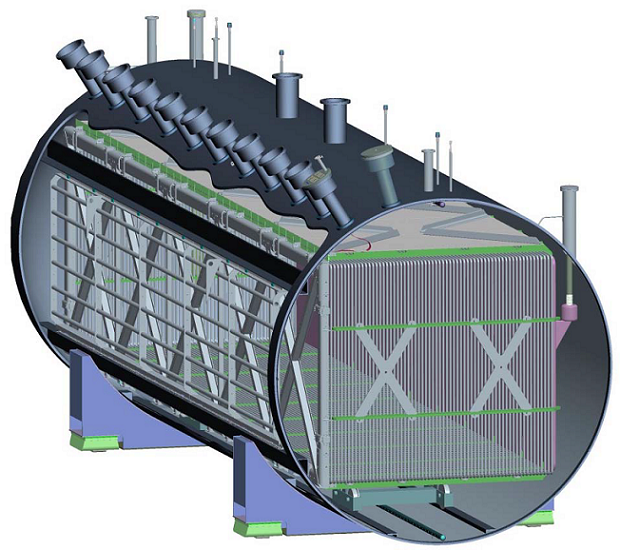}
\caption{The MicroBooNE LArTPC within the cylindrical cryostat, with the row of signal feedthrough chimneys and various other access ports also shown. Taken from \cite{MicroBooNE_TDR}.}
\label{fig:MicroBooNE_Cryostat}
\end{figure}

\begin{figure}[htp]
\centering
\includegraphics[width=0.52\textwidth]{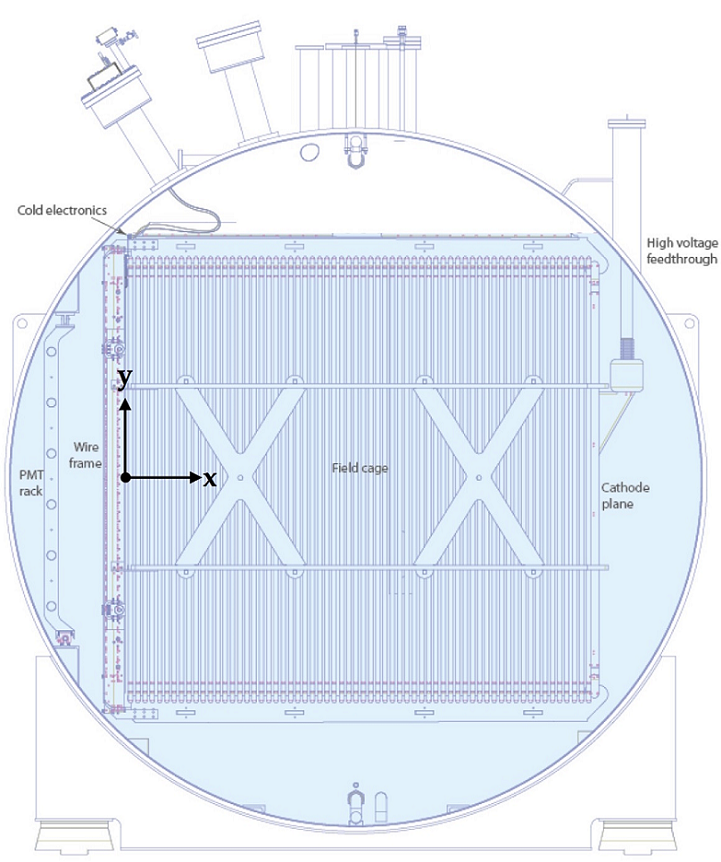}
\caption{A cross-section schematic depiction of the MicroBooNE detector, with some key components labelled. Also depicted is the coordinate system used by the experiment (the $z$ axis is directed towards the reader). Taken from \cite{MicroBooNE_TDR}.}
\label{fig:MicroBooNE_Detector}
\end{figure}

The design of MicroBooNE, shown in Figures~\ref{fig:MicroBooNE_Cryostat} and \ref{fig:MicroBooNE_Detector}, is similar to that of the individual ICARUS T600 drift volumes. The 86~tonnes of active LAr are contained in a single 2.325 $\times$ 2.560 $\times$ 10.368~m TPC \cite{MicroBooNE_TDR}, which is itself enclosed by a single-walled cylindrical cryostat. A 41~cm thick layer of foam insulation covers the outside of the cryostat, acting to reduce heat input from the detector's surroundings and ensure that the interior remains at cryogenic temperature. Plastic scintillator-based cosmic ray tagger planes are positioned outside the cryostat, parallel to the four long sides of the active volume \cite{MicroBooNE_CRT}.

\begin{figure}[hbp]
\vspace{5mm}
\centering
\includegraphics[width=0.52\textwidth]{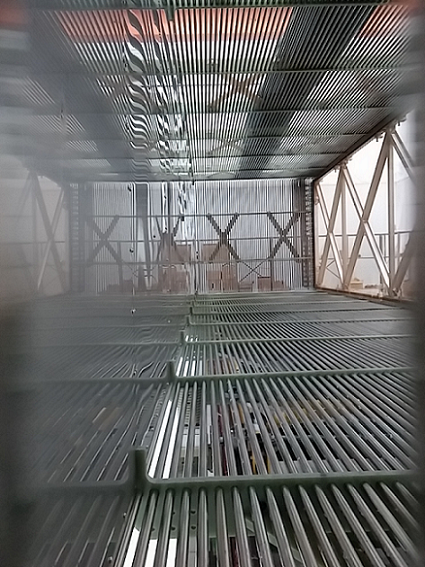}
\caption{An interior view of the MicroBooNE TPC, showing the cathode (left), the field cage loops (centre) and the support structure for the wire planes (right). Note that the cathode panels are highly reflective due to being polished for flatness. Taken from \cite{MicroBooNE_TDR}.}
\label{fig:MicroBooNE_TPC}
\end{figure}

A view of the interior of the MicroBooNE TPC is shown in Figure~\ref{fig:MicroBooNE_TPC}. The cathode is made up of 9 separate stainless steel panels, fixed to a supporting frame, and the bias is provided by a HV feedthrough of similar design to that used by ICARUS T600 \cite{MicroBooNE_TDR}. The drift volume is physically defined by 64 field cage loops, with 4~cm centre-to-centre spacing and connected to each other using a resistor chain to generate the drift field. To achieve the purity required for a 2.5~m electron drift distance, a dedicated LAr cryogenics and purification system operates in conjunction with the main MicroBooNE detector. This system - which is extensively described in \cite{MicroBooNE_TDR} - has been shown to reduce the levels of O$_2$ and H$_2$O contamination in the active LAr volume to the order of 100~ppt following a three-stage, multi-week purging, cooling and filling process.

\begin{figure}[h]
\vspace{2mm}
\centering
\includegraphics[width=0.85\textwidth]{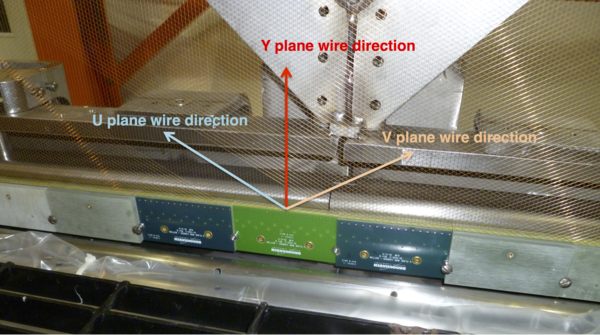}
\caption{A close-up view of the wire planes used in the MicroBooNE LArTPC. The plane labelling is as follows: ``U'' denotes the innermost (induction) plane, ``V'' is the middle plane (also induction), and ``Y'' is the collection plane. Taken from \cite{MicroBooNE_Wires}.}
\label{fig:MicroBooNE_Wires}
\end{figure}

Similarly to ICARUS T600, MicroBooNE utilises an APA consisting of 3 parallel wire planes for charge readout \cite{MicroBooNE_TDR}, with the two inner planes closest to the cathode operated in induction mode (with 2400 wires each) and the outermost one acting as a collection plane (with 3456 wires). The planes are spaced apart by 3~mm, as are the individual wires, and the induction wires are orientated at +60\textdegree{} (innermost plane) and -60\textdegree{} (middle plane) to the vertical. The wires on the collection plane are orientated vertically. A close-up view of the wire planes is shown in Figure~\ref{fig:MicroBooNE_Wires}. During standard operation of the detector, the planes are biased at -200, 0 and +440~V (innermost to outermost), which - along with the nominal cathode bias of -128~kV \cite{MicroBooNE_TDR} - gives a drift field of 500~V/cm and a maximum drift time of 1.6~ms.

\begin{figure}[ht]
\centering
\includegraphics[width=0.82\textwidth]{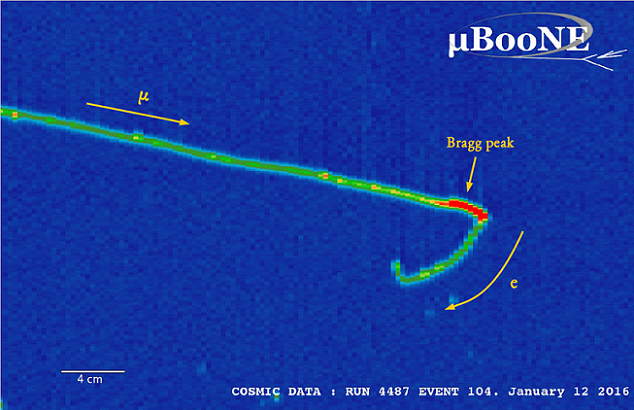}
\caption{A stopping muon and associated Michel electron candidate as recorded by the MicroBooNE collection plane. Taken from~\cite{MicroBooNE_MichelElectron}.}
\label{fig:MicroBooNE_StoppingMuon}
\end{figure}

Readout of the wire signals is accomplished through the use of ``application-specific integrated circuits'' (ASICs) that combine pre-amplifiers, digitisers, signal shaping and multiplexing into front-end assemblies arranged in a system broadly similar in function to that used in ICARUS T600 \cite{MicroBooNE_TDR}. However, one key difference is that MicroBooNE's front-end electronics are mounted directly on the APA support frames - i.e. within the LAr environment. This has necessitated the development of ``cold electronics'' with low noise and power consumption (thereby minimising heat output), but in return, the short distance and cryogenic environment between the wires and electronics significantly reduces the readout noise, therefore increasing the signal-to-noise ratio \cite{MicroBooNE_TDR}. (Additional general advantages of using cold front-end electronics over off-detector systems are presented in \cite{SBND_ReadoutElecs}.) Custom ``cold cables'' carry signals from the front-end electronics out of the cryostat via the signal feedthrough chimneys. On the external side of these chimneys are mounted the intermediate electronics, which include amplifiers that allow the signals to be carried to the off-detector readout and DAQ systems situated 20~m away from the cryostat \cite{MicroBooNE_TDR}. A stopping muon and associated Michel electron candidate, as recorded by the MicroBooNE collection plane, is pictured in Figure~\ref{fig:MicroBooNE_StoppingMuon}.

\begin{figure}[ht]
\centering
\includegraphics[width=0.82\textwidth]{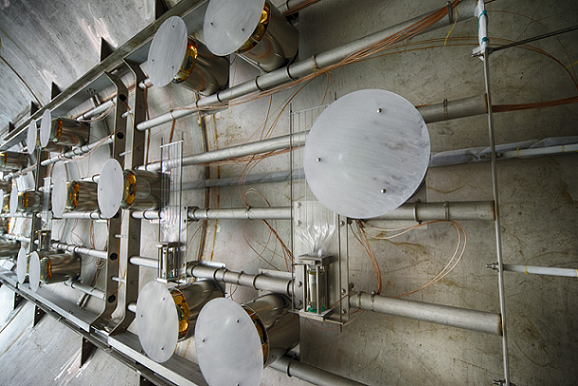}
\caption{The MicroBooNE light collection system mounted on the interior of the cryostat. The two subsystems can be seen: the primary array of 32 8-inch diameter Hamamatsu R5912-02MOD PMTs with attached TPB-coated acrylic disks, and the secondary set of 4 (vertically orientated) lightguide paddles. Taken from \cite{MicroBooNE_TDR}.}
\label{fig:MicroBooNE_PMTs}
\end{figure}

The light collection system in MicroBooNE serves three separate purposes \cite{MicroBooNE_TDR}. The first is, as in ICARUS T600, to provide an event start time by observing prompt scintillation light. The second use is to act as part of the detector's dedicated triggering mechanism for beam neutrino interactions that produce protons with kinetic energy of $\geq$ 40~MeV. (Such interactions also produce a distinct pulse of scintillation light that will be correlated with the known time of a beam spill passing through the detector.) The final purpose of the system is to directly observe the light emitted by lower energy (5 - 10~MeV) electrons produced in interactions between LAr and supernova neutrinos. The complete light collection system is shown in Figure~\ref{fig:MicroBooNE_PMTs}, and comprises two subsystems \cite{MicroBooNE_TDR}. The first is an array of 32 8-inch diameter Hamamatsu R5912-02MOD PMTs, with a TPB-coated acrylic disk individually mounted in front of each one. This separation of the wavelength shifter from the PMT differs from the direct coating method demonstrated at ICARUS T600 (and used in a number of other detectors), but has been done at MicroBooNE for purposes of ease-of-installation and improving the quality of the TPB in the cryogenic environment \cite{MicroBooNE_TDR}. Electromagnetic isolation is achieved by housing each PMT in a cryogenic mu metal shield. The second light collection subsystem consists of 4 50.8~cm long lightguide paddles, built up of 6 bent acrylic bars (coated with wavelength shifter) that guide incident light to a coupled PMT. This secondary system has been installed in MicroBooNE primarily for R\&D purposes: each paddle has a larger coverage than a single PMT, while occupying significantly less space due to their thin shape, making such devices important for future large-scale detectors, as they could allow a correspondingly large-scale light collection system to be realised for reduced complexity, cost and installation time.

Alongside development, characterisation and operation of the detector hardware, MicroBooNE's half-decade of operation has allowed for a strong analysis program to be undertaken, covering a wide range of topics including (but by no means limited to) single particle and EM shower reconstruction, energy response calibration and resolution measurements, background identification and rejection techniques, vertex identification and reconstruction, and the use of neural networks for event reconstruction. A comprehensive list of published results can be found in \cite{MicroBooNE_Papers}.

\subsection{SBND}
\label{subsec:SBND}
The SBND (``\textbf{S}hort-\textbf{B}aseline \textbf{N}ear \textbf{D}etector'') is the newest LArTPC experiment at Fermilab. As its name implies, it is intended to be the near detector of the SBN Program on Fermilab's Booster Neutrino Beamline, and as such it will receive the full unoscillated contents of the beam - more than 6.5 million $\nu_{\mu}$ events and 55,000 $\nu_{e}$ events over three years of operation \cite{SBND_Overview}. Such high statistics will allow SBND to perform high precision measurements across a wide range of topics in Neutrino Physics - including LAr interaction cross sections, the search for light and heavy sterile neutrinos, and observation of rare and non-standard interactions.

\begin{figure}[htp]
\centering
\includegraphics[width=0.65\textwidth]{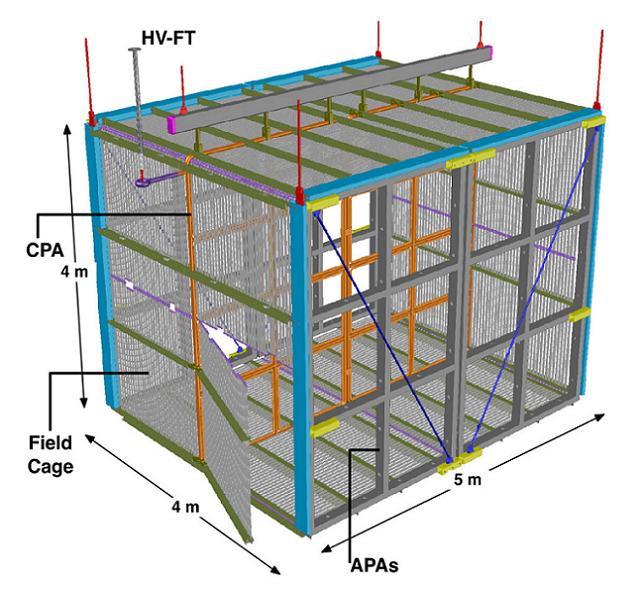}
\caption{The SBND LArTPC structure, with some of the major components labelled: the cathode plane assembly (CPA), the wire-based anode plane assemblies (APAs), the field cage and the HV-FT (high voltage feedthrough). Taken from \cite{SBND_Images}.}
\label{fig:SBND_Detector}
\end{figure}

\begin{figure}[hbp]
\vspace{5mm}
\centering
\includegraphics[width=0.50\textwidth]{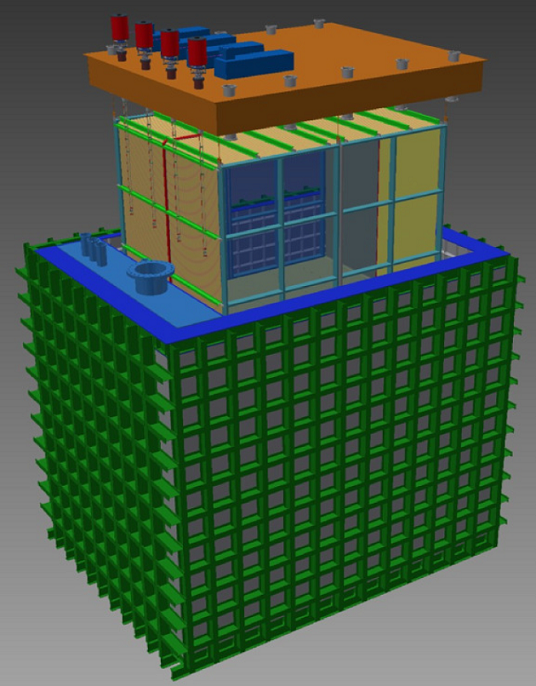}
\caption{Conceptual design of the SBND membrane cryostat (green), with the TPC structure suspended from the removable lid (orange). Taken from \cite{SBND_Images}.}
\label{fig:SBND_Cryostat}
\end{figure}

The SBND LArTPC structure is shown schematically in Figure~\ref{fig:SBND_Detector}. Sharing a similar design to the two ICARUS T600 modules, the 4 $\times$ 4 $\times$ 5~m detector contains an active LAr mass of 112~tonnes, and features a vertically-orientated ``cathode plane assembly'' (CPA) with two independent drift volumes positioned back-to-back and wire-based APAs located on the outer sides. The various elements of the photon detection system \cite{SBND_PhotonDetection}, which is once again used to directly observe the prompt scintillation light, are mounted behind the wires of the APAs. The entire LArTPC structure is suspended from the lid of the SBND membrane cryostat, as shown in Figure~\ref{fig:SBND_Cryostat}, and cosmic ray taggers comprised of multiple planes of scintillator bars surround the structure \cite{SBN_Review}. The use of a membrane cryostat instead of a more standard vacuum vessel allows SBND to act as a prototype and testing platform for the expected use of membrane cryostats by the (Proto)DUNE single- and dual-phase detectors. 

The SBND CPA consists of 16 stainless steel mesh panels held in a stainless steel support frame \cite{SBN_Review}. The cathode bias - nominally of -100~kV - is transferred via a high voltage feedthrough of similar design to that used by ICARUS T600, which connects to the CPA at a contact point made between the feedthrough's spring-loaded tip and a torus that is welded to the support frame.  Rather than using continuous electrode loops like those employed on ICARUS T600 and MicroBooNE, the field-shaping in the SBND drift volumes is performed by a series of roll-formed aluminium profiles \cite{SBN_Review} mounted onto a structural framework positioned around the perimeter of the CPA, as shown in Figure~\ref{fig:SBND_Detector}. This field cage design is the same as that of the ProtoDUNE detectors, and will also be used in both the single- and dual-phase DUNE LArTPCs as well. The nominal drift field in SBND is 500~V/cm, giving a maximum drift time of 1.3~ms in each of the 2~m wide drift volumes \cite{SBN_Review}.

\begin{figure}[h]
\vspace{2mm}
\centering
\includegraphics[height=0.20\textheight]{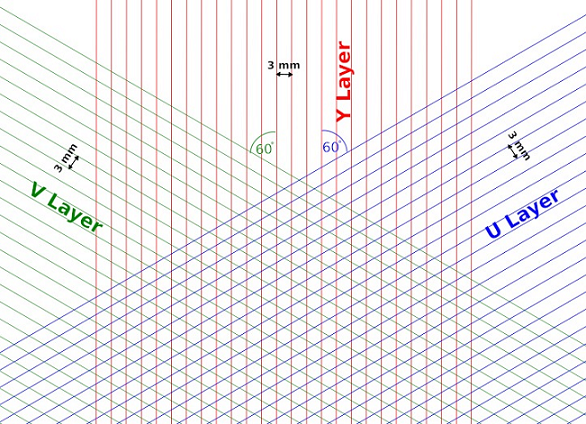}
\hspace{1mm}
\includegraphics[height=0.20\textheight]{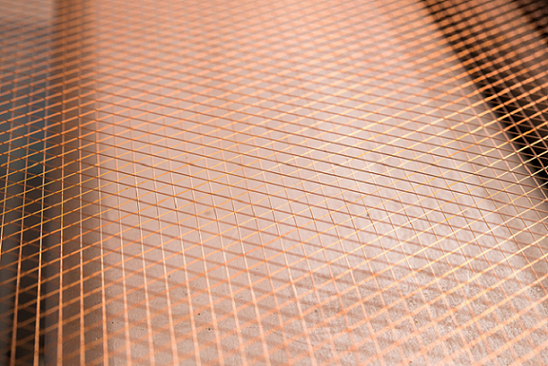}
\caption{(Left) A schematic layout of the wire planes in SBND, showing the U (innermost) and V (middle) induction planes with wires at +60\textdegree{} and -60\textdegree{} to vertical respectively, and the Y (outermost) collection plane with wires orientated vertically. (Right) The wire planes installed on one of the SBND APAs. Both taken from \cite{SBND_WirePlanes}.}
\label{fig:SBND_Wires}
\end{figure}

In terms of the fundamental design, each of the SBND APAs \cite{SBND_WirePlanes} is identical to that of MicroBooNE: 3 parallel planes of wires denoted as U, V and Y in order of innermost to outermost, respectively orientated at angles of +60\textdegree{}, -60\textdegree{} and 0\textdegree{} in relation to vertical, and with both wire-to-wire and inter-plane spacings of 3~mm. This is depicted schematically in Figure~\ref{fig:SBND_Wires} (left) and photographed in Figure~\ref{fig:SBND_Wires} (right). The planes are biased at -200~V, 0~V and +430~V (U, V and Y respectively) \cite{SBND_WirePlanes}, once again resulting in two induction planes and one collection plane. The wires in a single APA are installed in a stainless steel frame that forms a 2 $\times$ 3 grid of ``windows'', and so each of the drift volumes' outer faces therefore consists of two such APAs, as previously shown in Figure~\ref{fig:SBND_Detector}. The APAs on each face are manufactured and wired independently, before being electronically coupled so that they effectively function as a single structure. To ensure flatness of the wires across each APA's full height and width and to support them away from their soldered mounting points on the perimeter of the frames, specially designed combs are installed on the frames' cross-beams \cite{SBND_WirePlanes}. As with the use of a membrane cryostat, the design, manufacturing and testing of the SBND APAs has acted as prototyping for the DUNE experiments, which are expected to use wire planes of similar design and size.

Like MicroBooNE, SBND utilises a combination of cold and warm electronics in its readout system \cite{SBND_ReadoutElecs}. Initial signal collection, amplification and digitisation is performed by ASICs located on cold front-end motherboards that are mounted on the APAs within the LAr environment, with 1 motherboard collating signals from 128 wires. The design of these motherboards is in part based on similar cold electronics devices developed specifically for the single-phase ProtoDUNE detector, but commercial off-the-shelf components are also used in SBND \cite{SBND_ReadoutElecs}. The signals from the front-end motherboards are carried via the cryostat's four signal feedthrough flanges to ``warm interface electronics crates'' (WIECs), one of which is mounted directly to the external side of each feedthrough. These crates collate the signals from up to 24 front-end motherboards (3072 wires) before passing them to the off-detector back-end acquisition and recording systems.

\begin{figure}[h]
\vspace{2mm}
\centering
\includegraphics[height=0.30\textheight]{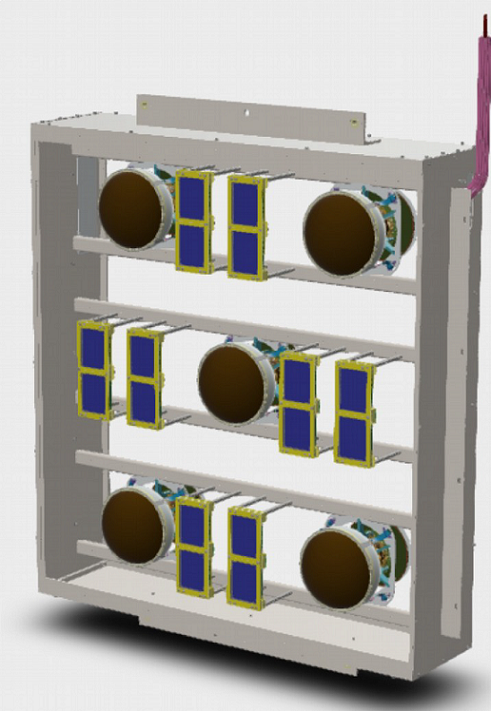}
\caption{A model of one of the SBND ``PDS box'' assemblies, containing 5 8-inch diameter Hamamatsu PMTs and 8 X-ARAPUCA (``Argon R\&D Advanced Program at UniCAmp'') devices. The central PMT is uncoated, and the rest are coated with TPB. These assemblies are installed behind the wire planes of the APAs. Depending on the exact location, some of the X-ARAPUCAs are replaced by ARAPUCAs, and/or coated with TPB \cite{SBND_PhotonDetection}.  Taken from \cite{SBND_Overview}.}
\label{fig:SBND_PDSBox}
\end{figure}

The SBND ``photon detection system'' (PDS) \cite{SBND_PhotonDetection} also offers an opportunity for R\&D and prototyping of technologies and designs that could prove important for future large-scale LArTPCs. The basis of the PDS is a system of 120 8-inch diameter Hamamatsu PMTs (60 per drift volume), 96 of which are coated with TPB - making them sensitive to both VUV and visible wavelengths. These operate in a similar way to the PMT assemblies previously described at ICARUS T600 and MicroBooNE: detecting the prompt scintillation light for the purpose of event triggering. The remainder of the PMTs are uncoated, and are therefore only sensitive to direct visible wavelength photons. On their own, these uncoated PMTs therefore do not directly observe any VUV scintillation light, but a second component of the PDS is a total of 38m$^2$ of TPB-coated reflective foil mounted on both sides of the CPA \cite{SBND_PhotonDetection}. This wavelength-shifts any VUV scintillation light that is emitted towards the CPA, and then reflects it at visible wavelengths in the direction of the PMTs, allowing it to be detected and therefore significantly increasing the per-event light yield compared to using only the TPB-coated PMTs. In addition to the PMTs, two relatively new devices called ARAPUCA and X-ARAPUCA are being tested as part of the SBND PDS \cite{SBND_PhotonDetection}. The ARAPUCA (``Argon R\&D Advanced Program at UniCAmp'') device \cite{ARAPUCA} is a form of light trap consisting of a box with a highly reflective interior, and an aperture that combines a dichroic filter and two wavelength shifters - one on the external surface that shifts incident light so as to allow it through the filter, and the other on the interior that shifts it into the opaque range of the filter, thus trapping it inside. The box is itself directly coupled to a photon-sensitive sensor (such as a SiPM), and detection of the light relies on the photons reflecting at the correct angle to be incident on the sensor before they are absorbed. The X-ARAPUCA \cite{XARAPUCA} is a direct advancement of the ARAPUCA, and contains an internal acrylic slab with embedded wavelength shifter, to which the photo-sensitive sensor is coupled to. This slab serves the same purpose as the interior wavelength shifter of the ARAPUCA, but also acts as a lightguide between the device's interior and the sensor - thereby significantly increasing the detection efficiency of this device over the original ARAPUCA. As with the PMTs, some of the (X-)ARAPUCAs are coated with TPB, and the rest are not. The PMTs and (X-)ARAPUCA devices are installed in ``PDS box'' assemblies that are mounted behind the APA wire planes, fitting into the APA frame windows. One such assembly is shown in Figure~\ref{fig:SBND_PDSBox}.

At the time of writing, SBND's TPC components are being assembled in the Fermilab D0 Building. After assembly is complete, the entire structure will be moved to its cryostat located in the dedicated SBND Building.

\subsection{LArIAT}
Due to their location and operation on a neutrino beam, the three LArTPCs in the SBN Program are designed to observe the signal from such a beam - that is, neutrino interactions. However, it is equally important to be able to identify and characterise the backgrounds to that signal: non-neutrino particle interactions that produce the same, or similar, detectable products. For example, charged-current $\nu_{e}$ elastic scattering results in an electron-induced electromagnetic shower, but a gamma-induced shower can be produced by the decay of a $\pi^{0}$ into a $\gamma$ pair followed by one of the $\gamma$'s undergoing pair production. LArTPCs are potentially capable of distinguishing between electron- and gamma-induced showers through reconstruction of any pre-vertex tracks (which should be present in electron-induced showers, but not in gamma-induced ones), and measurement of the total deposited energy (which should be twice as high for the gamma-induced $e^{+}e^{-}$ pair than for the single $e^{-}$ that initiates the electron-induced shower). The ArgoNeuT experiment produced the first (and currently only) exploration of shower separation in a LArTPC, using a small dataset \cite{ArgoNeuT}. The LArIAT (``\textbf{L}iquid \textbf{Ar}gon \textbf{I}n \textbf{A} \textbf{T}estbeam'') experiment \cite{LArIAT_TDR}, located at the Fermilab Test Beam Facility, aims to further that investigation using higher statistics, as well as perform precision measurements of other charged particle LAr interactions - not just to better understand their role as background to neutrino interactions, but also because they are naturally scientifically interesting in their own right.

\begin{figure}[h]
\vspace{2mm}
\centering
\includegraphics[height=0.38\textheight]{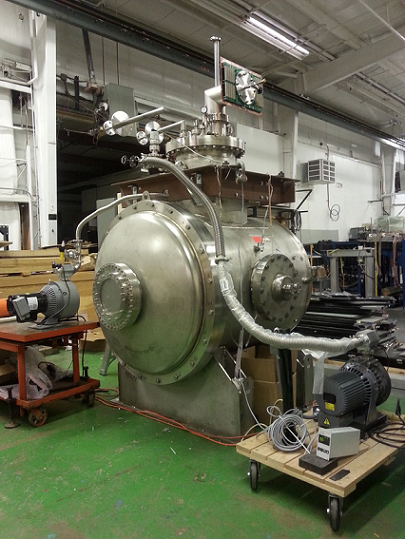}
\hspace{1mm}
\includegraphics[height=0.38\textheight]{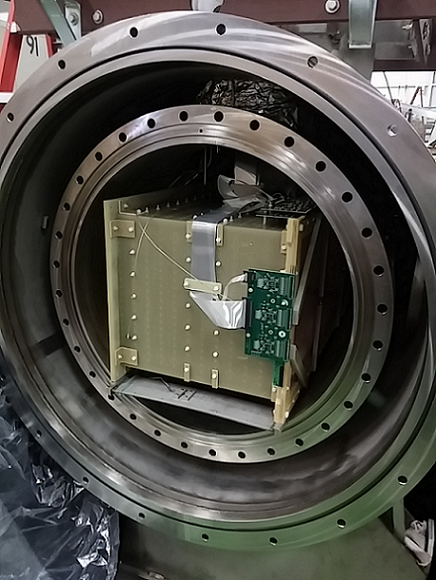}
\caption{(Left) The LArIAT cryostat, prior to relocation into the Fermilab Test Beam Facility. The beam window on the outer upstream end-cap can be seen, as well as the flange connected to the PDS located on the cylindrical side, and various other off-detector hardware such as scroll pumps for the vacuum jacket and components of the LAr recirculation system. (Right) The interior of the cryostat, showing the position of the TPC within the inner vessel, and the cold front-end electronics boards connected to the wire-based APA. Both taken from \cite{LArIAT_TDR}.}
\label{fig:LArIAT_Cryostat}
\end{figure}

LArIAT operated with various hardware configurations between 2015 and 2017 \cite{LArIAT_TDR}, but the basic detector design remained the same throughout. The cryostat, which was inherited from ArgoNeuT and is shown in Figure~\ref{fig:LArIAT_Cryostat}, is a cylindrical double-walled design, with the space between the inner and outer walls acting as a vacuum jacket for heat insulation. The TPC is situated within the inner vessel. The two ends of the cryostat, which are referred to as ``upstream'' (the end through which the beam enters the detector) and ``downstream'' (through which the beam exits), are sealed by inner and outer convex end-caps. A thin titanium beam window has been added to the outer upstream end-cap, and the inner upstream end-cap has been modified with the addition of a hollow volume referred to as the ``excluder'' \cite{LArIAT_TDR}. Together, the beam window and excluder significantly reduce the amount of uninstrumented material that the beam encounters before entering the active LAr within the drift volume.

The LAr purification and recirculation system at LArIAT is based on the system developed at the Liquid Argon Purity Demonstrator \cite{LiquidArgonPurityDemo}. That facility demonstrated that it is possible to achieve extremely high LAr purity, and therefore long electron lifetimes (exceeding 6~ms), without first having to evacuate the containing vessel to vacuum. This is particularly important for detectors which use membrane cryostats (such as the previously discussed SBND, as well as the ProtoDUNE detectors and the future DUNE LArTPCs), as this type of vessel is not capable of holding an internal vacuum. The LArIAT system is capable of reducing impurities to the level of 100~ppt \cite{LArIAT_TDR}, resulting in electron drift times of several milliseconds.

\begin{figure}[h]
\vspace{2mm}
\centering
\includegraphics[width=0.60\textwidth]{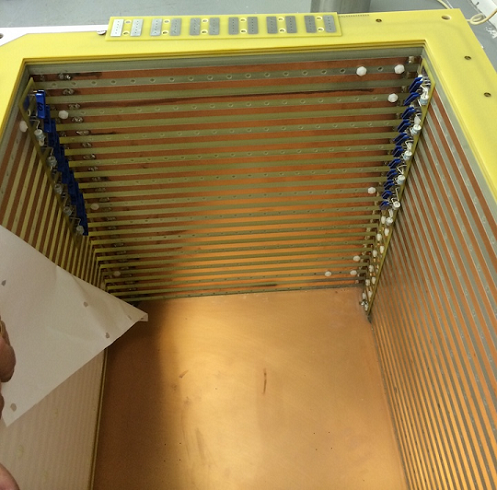}
\caption{A view of the inside of the LArIAT field cage, showing the solid copper plane that was initially used for the cathode, and the copper strips that make up the field cage attached on the outer walls of the TPC. A TPB-coated reflector foil can be seen partially attached to the left-side wall. The resistor chain for regulating the electric field is attached to the right-side wall near the corner. Taken from \cite{LArIAT_TDR}.}
\label{fig:LArIAT_FieldCage}
\end{figure}

The LArIAT drift volume is rectangular, with dimensions of 47 $\times$ 40 $\times$ 90~cm \cite{LArIAT_TDR}, giving an active LAr mass of 237~kg. The cathode is positioned on the left side of the TPC when viewed from the upstream end. Initially, the cathode consisted of a single copper plate, but towards the end of operation this was replaced by a stainless steel mesh - acting as a prototype for the cathode that has now been installed in SBND. The cathode bias is provided by a HV feedthrough of similar design to that used by ICARUS T600, and was set at -23.5kV \cite{LArIAT_TDR}. The outer walls of the TPC also double as the mounting points for the field-shaping elements - constructed out of 1~cm wide copper strips with a 1~cm centre-to-centre spacing, as shown in Figure~\ref{fig:LArIAT_FieldCage}. As with previously discussed field cages, the drift field is created through the use of a resistor chain attached to the field cage elements. All four sides of the field cage are covered by TPB-coated reflective foils, to wavelength-shift the VUV scintillation light to visible wavelengths that can be observed by the LArIAT PDS (described below). The same foils were also mounted onto the cathode \cite{LArIAT_TDR}.

The APA of LArIAT consists of three wire planes, although only two of them were actually used for charge readout. The closest plane to the cathode is the ``shield'' plane - comprised of wires that are vertically orientated, but not read out. Moving outwards, this is followed by one induction plane and the collection plane, with wires orientated at +60\textdegree{} and -60\textdegree{} to the vertical respectively \cite{LArIAT_TDR}. (From a purely geometric standpoint, two 2D views is indeed enough to fully reconstruct the three-dimensional particle track, but the addition of a third view, as in other experiments' wire-based charge readouts, greatly increases the reconstruction performance and precision.) The wire spacing in all three planes was set at 4~mm during much of LArIAT's operation, with the biases being -298 (shield), -18.5 (induction) and +338~V (collection). Towards the end of the data-taking period, the spacing was changed to 5 and then 3~mm, necessitating corresponding changes in the biases in order to ensure that 100\% of the drifted electrons still pass through both the shield and induction planes \cite{LArIAT_TDR}. These alternate setups were used to test the effect of wire spacing on the overall performance of the experiment, so as to inform future LArTPC design.

The signals from the 480 wires are first passed to ten front-end electronics boards mounted on the side of the TPC, i.e. within the cryogenic environment. The ASICs on these boards perform initial amplification and digitisation of the signals (as well as providing bias to the individual wires) \cite{LArIAT_TDR}, before they are sent through the single feedthrough on the top of the cryostat to the warm electronics mounted on the external side of the flange. Additional amplification and noise cancellation is performed here, and the signals are then carried to the array of digitisers situated 8~m from the detector vessel \cite{LArIAT_TDR}. This distance is required to ensure that charged particles from the beam do not pass through the electronics, which would otherwise result in considerable noise being introduced.

\begin{figure}[h]
\vspace{2mm}
\centering
\includegraphics[width=0.60\textwidth]{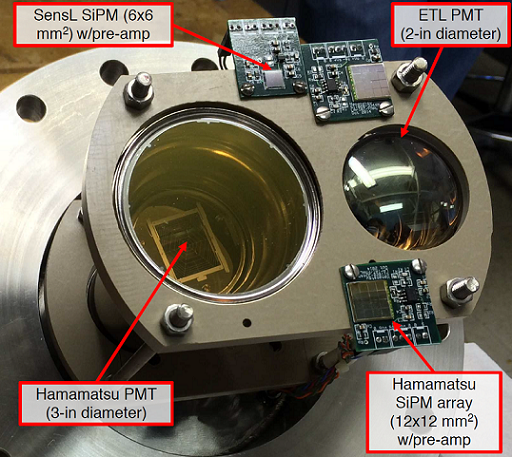}
\caption{The LArIAT PDS, showing the two PMTs (3 and 2-inch diameter) and the two silicon photomultipliers (SiPMs). Taken from \cite{LArIAT_TDR}.}
\label{fig:LArIAT_PDS}
\end{figure}

As with the wire planes, the LArIAT PDS - shown in Figure~\ref{fig:LArIAT_PDS} - was used not just for data collection, but also to evaluate various hardware and design configurations throughout the experiment's operation time. Along with the aforementioned TPB-coated reflective foils mounted on the field cage, a variety of photon detectors are positioned approximately 5~cm behind the collection plane, on a PEEK support structure that is itself mounted onto a flange that passes through both walls of the cryostat \cite{LArIAT_TDR}. This allows for the removal and replacement of the PDS without needing to open the rest of the detector. Initially, the PDS consisted of two PMTs with 3 and 2-inch diameters, but these were augmented by two silicon photomultipliers at various points, as well as an ARAPUCA \cite{LArIAT_TDR}. In addition, the 2-inch PMT and one of the photomultipliers were given a TPB coating, making them sensitive to scintillation light arriving directly from the point of emission rather than only the visible light that had been reflected from the foils.

\begin{figure}[h]
\centering
\includegraphics[width=0.80\textwidth]{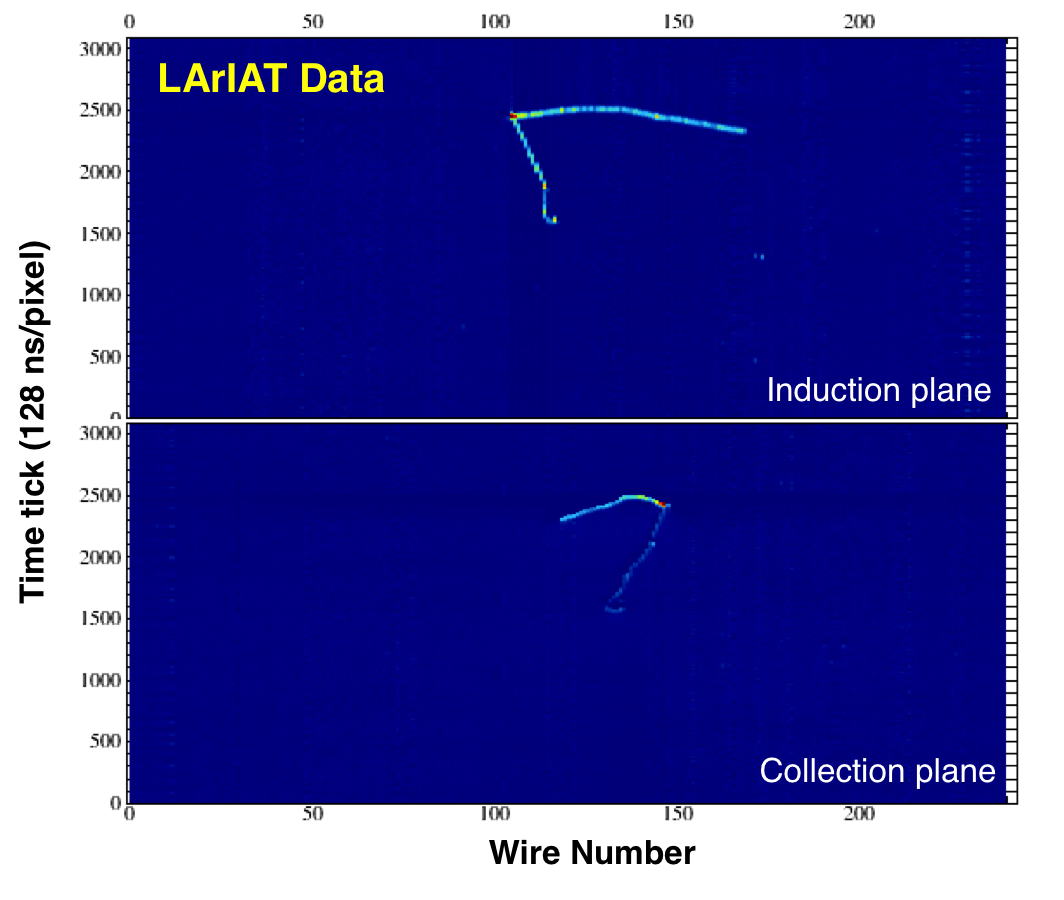}
\caption{Two-dimensional views of a stopping muon and associated Michel electron, as recorded by the LArIAT induction and collection wire planes. Taken from \cite{LArIAT_MichelElecs}.}
\label{fig:LArIAT_StoppingMuon}
\end{figure}

The high statistics of data collected by LArIAT during its operation has allowed not only high precision characterisation of the detector itself (thereby providing valuable insights into the hardware and designs that were used), but also successful investigations of a number of charged particle LAr interactions as originally envisioned. These include the world first measurement of the $\pi$-Ar cross-section \cite{LArIAT_PiCrossSec}, high resolution calorimetric reconstruction of low-energy electrons \cite{LArIAT_Electrons} and identification and analysis of Michel electrons using only scintillation light-based triggering \cite{LArIAT_MichelElecs}. An example of an identified stopping muon and its associated Michel electron, as seen by the induction and collection wire planes, is shown in Figure~\ref{fig:LArIAT_StoppingMuon}. The LArIAT detector itself is currently no longer operating, but analysis of the data continues, and a number of its findings relating to LArTPC design and hardware have already proved important for the latest generation of detectors.

\subsection{DUNE (Single-Phase Far Detector)}
\label{subsec:DUNE_SP}
The culmination of over forty years of worldwide research and development of LArTPC technology is DUNE (``\textbf{D}eep \textbf{U}nderground \textbf{N}eutrino \textbf{E}xperiment'') \cite{DUNE_IDRv1}. This project is a long-baseline experiment utilising a neutrino beam produced at Fermilab, a near detector complex located 575~m from the beam target, and a far detector complex 1300~km away and 1.5~km underground at the Sanford Underground Research Facility (SURF) in South Dakota. The near detector is expected to implement a variety of technologies \cite{DUNE_NearDetector}, including a LArTPC based on the in-development ArgonCube modular TPC and pixelated charge readout systems (described in Section~\ref{subsec:ArgonCube} below).

\begin{figure}[h]
\vspace{2mm}
\centering
\includegraphics[width=0.95\textwidth]{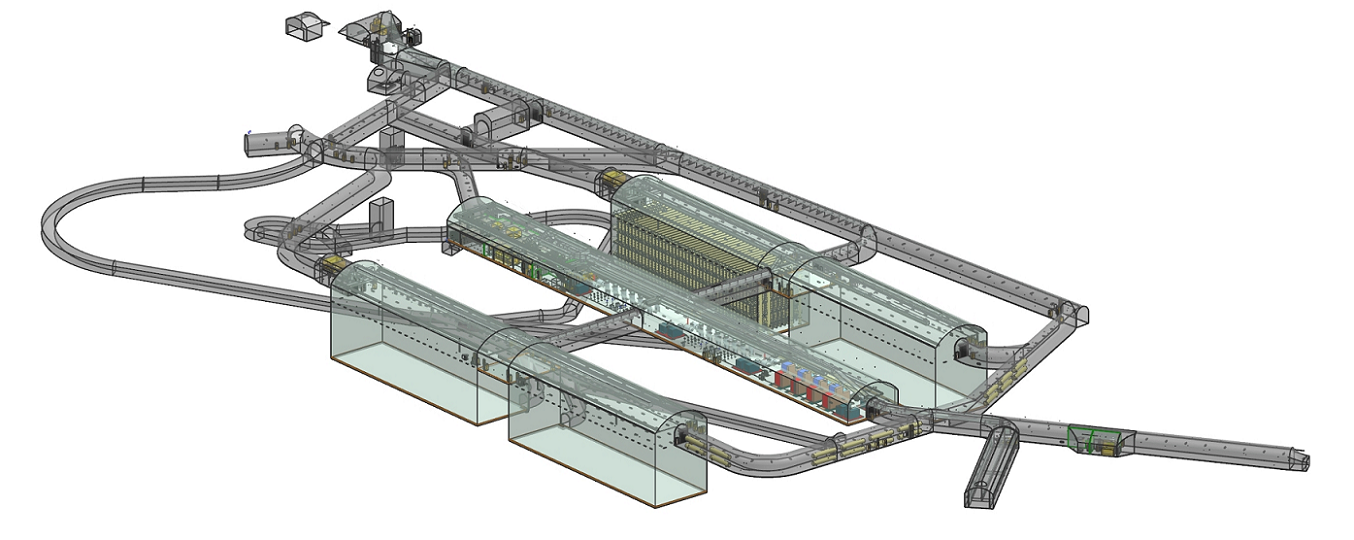}
\caption{The proposed layout of the Sanford Underground Research Facility, which will host the DUNE far detector complex approximately 1.5~km below ground level. The facility will consist of four large caverns for housing the LArTPCs, a number of smaller spaces for various aspects of operation such as control rooms, electronics and cryogenic systems, and connecting access tunnels. Taken from \cite{DUNE_IDRv1}.}
\label{fig:DUNE_SURFComplex}
\end{figure}

The DUNE far detector complex at SURF is shown in Figure~\ref{fig:DUNE_SURFComplex}. At the heart of the facility will be four LArTPCs (referred to as ``modules''), each measuring 14.0 $\times$ 14.1 $\times$ 62.0~m and containing approximately 17,500~tonnes of LAr \cite{DUNE_IDRv1}. The first module to be constructed will use the standard single-phase design (referred to hereafter as ``DUNE-SP'', and described in this section). Moreover, the construction of a dual-phase module (``DUNE-DP'', described in Section~\ref{subsec:DUNE_DP}) is also planned. The technology to be implemented in the remaining two modules is yet to be decided. It must be noted that, while the Physics, engineering and operational requirements of the DUNE modules are well understood, their designs are still technically conceptual. Ongoing development, testing and refinement of these designs is being carried out primarily at the single- and dual-phase ProtoDUNE experiments (commonly known as ``ProtoDUNE-SP'' and ``ProtoDUNE-DP'', and described in Sections~\ref{subsec:ProtoDUNE_SP} and \ref{subsec:ProtoDUNE_DP} respectively). These detectors use the same designs, technologies and manufacturing and assembly procedures to DUNE-SP and DUNE-DP, allowing all aspects of the experimental hardware and operation to be tested under effectively the same conditions as will be faced by DUNE.

\begin{figure}[ht]
\centering
\includegraphics[width=0.65\textwidth]{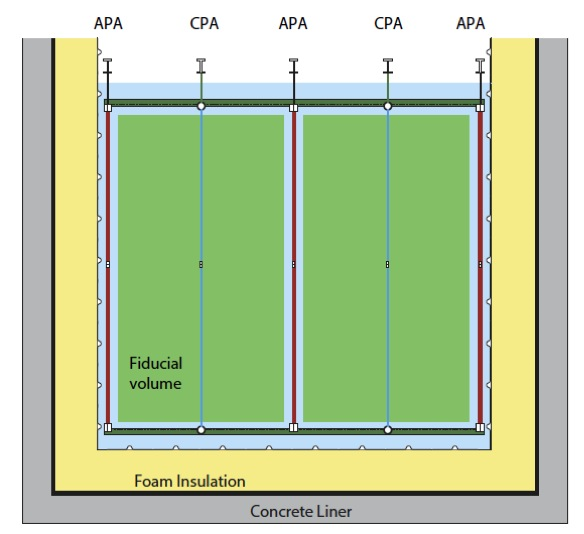}
\caption{A side-on view of the single-phase DUNE module, showing the layout of the CPAs and APAs in the liquid argon volume (blue), as well as the surrounding layers of the membrane cryostat. Taken from \cite{DUNE_IDRv2}.}
\label{fig:DUNE-SP_Detector}
\end{figure}

A side-on view of DUNE-SP is shown in Figure~\ref{fig:DUNE-SP_Detector}. Much like ICARUS T600, the module will be comprised of multiple drift volumes - in this case, four, with the two CPAs and the middle APA being shared by their respective adjacent volumes. While this is of course a well-established design, the considerable physical size of DUNE-SP means that certain considerations must be made for the individual components.

\begin{figure}[ht]
\centering
\includegraphics[width=0.45\textwidth]{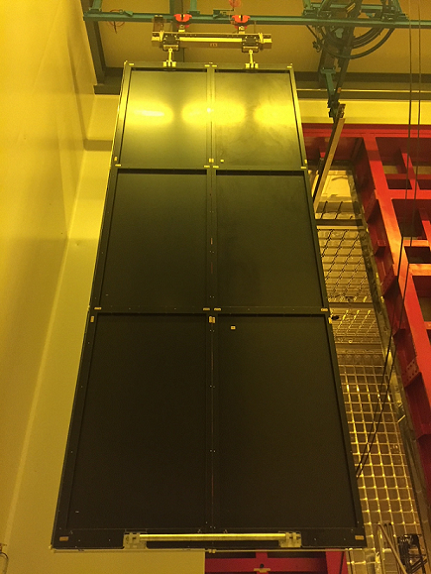}
\hspace{1mm}
\includegraphics[width=0.45\textwidth]{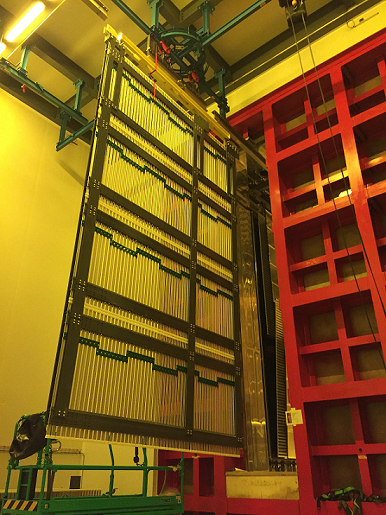}
\caption{(Left) A section of one of the ProtoDUNE-SP CPAs, showing 6 of the 1.2 $\times$ 2~m Kapton-laminated FR4 RP elements mounted in the FR4 supporting frame. Each complete DUNE-SP CPA will consist of 300 elements, i.e. 50 of these sections. (Right) One of the ProtoDUNE-SP field cage endwalls, using the same design that will be implemented in DUNE-SP. The individual rolled aluminium profiles can be seen, mounted on the FR4 structural framework and electrically interconnected. Similar assemblies are present on the top, bottom and other end of the TPC. Both taken from \cite{DUNE_IDRv2}.}
\label{fig:DUNE-SP_FieldCage}
\end{figure}

Rather than metal meshes or plates as used in smaller experiments previously described, each of the CPAs in DUNE-SP will consist of 3~mm thick FR4 with carbon-impregnated Kapton lamination on both sides \cite{DUNE_IDRv2}. This structure has been chosen based on consideration of the large total energy ($\approx$ 400~J) stored in the electric field created by such a large-area cathode. In the event of a discharge, the low surface resistivity of a metal cathode would allow all of this energy to be released within nanoseconds, inducing currents in the APA wires that could be high enough to cause damage in the readout electronics. However, Kapton has a high surface resistivity, which reduces the speed of energy dissipation enough to protect surrounding detector components \cite{DUNE_IDRv2}. To preserve the flatness of each CPA across its entire surface, the FR4 sheet of a single assembly is divided into 300 1.2 $\times$ 2~m ``RP elements'', mounted to a frame also constructed of FR4 (so that both frame and sheet experience the same amount of thermal contraction under cryogenic conditions). Figure~\ref{fig:DUNE-SP_FieldCage} (left) shows a section of one CPA, assembled for use in ProtoDUNE-SP but following the same design as will be used in DUNE-SP.

As previously noted in Section~\ref{subsec:SBND}, the DUNE-SP (and ProtoDUNE-SP) field cage follows the same design that is currently being implemented in SBND: a structural framework (which, in DUNE-SP, will be constructed from pultruded fiber-reinforced polymer beams) covering the top, bottom and end sides of each module, onto which are mounted rolled aluminium profiles \cite{DUNE_IDRv2}. This design has been chosen primarily for reasons of scale: it is impractical for a detector as large as DUNE-SP to utilise continuous field cage loops such as those on ICARUS T600 and MicroBooNE. A view of one of the ProtoDUNE-SP field cage endwalls is shown in Figure~\ref{fig:DUNE-SP_FieldCage} (right). The anticipated nominal cathode bias is -180~kV, provided through a dedicated HV feedthrough based on the venerable ICARUS T600 design, giving a drift field of 500~V/cm across each of the 3.6~m wide drift volumes.

\begin{figure}[h!]
\vspace{2mm}
\centering
\includegraphics[width=0.90\textwidth]{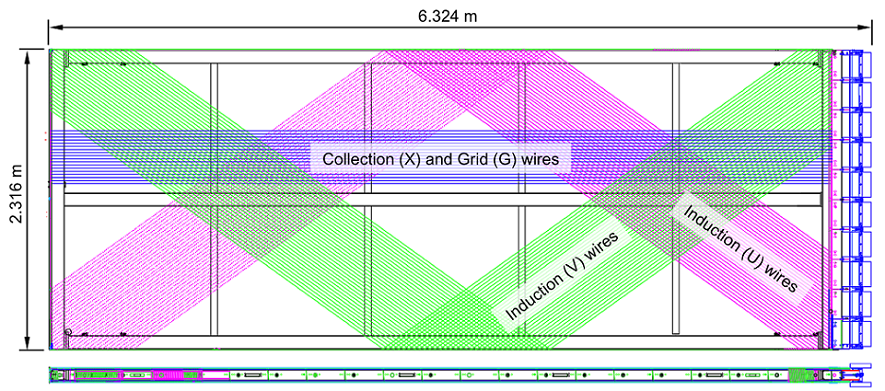}
\caption{A schematic view of one of the 6 $\times$ 2.3~m APAs to be used in DUNE-SP. The four wire planes (U, V, X and G) are labelled, with the wrapping of the induction plane wires also indicated. Elements of the cold front-end electronics are depicted by the blue boxes on the right side. Note that when installed in the detector, the electronics boxes will be at the top - i.e. the APA will be rotated anticlockwise 90\textdegree{} from this drawing. Taken from \cite{DUNE_IDRv2}.}
\label{fig:DUNE-SP_APASchematic}
\end{figure}

\begin{figure}[ht] 
\centering
\includegraphics[width=0.85\textwidth]{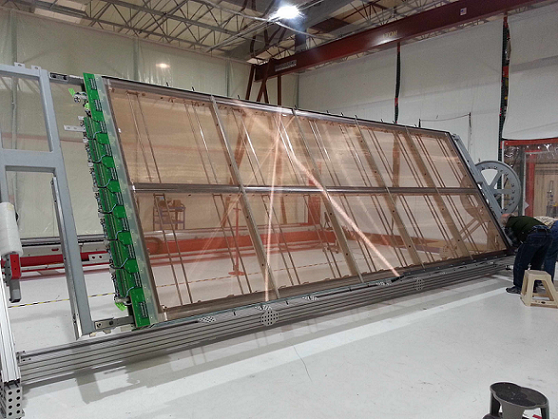}
\caption{A completed APA, with all wire planes in place and on-frame electronics installed. This particular APA is currently installed at ProtoDUNE-SP (see Section~\ref{subsec:ProtoDUNE_SP}), but the design is identical to that of DUNE-SP. Taken from \cite{DUNE_IDRv2}.}
\label{fig:DUNE-SP_APAImage}
\end{figure}

The wire-based APAs will follow the same three plane (2 induction + 1 collection) design utilised in other single-phase LArTPCs, but due to the large area they are required to cover, each ``module-sized APA'' will in fact consist of 50 individual but electrically connected 6 $\times$ 2.3~m APAs (with therefore a total of 150 such APAs in DUNE-SP) \cite{DUNE_IDRv2}. The schematic layout of one such APA is depicted in Figure~\ref{fig:DUNE-SP_APASchematic}, with Figure~\ref{fig:DUNE-SP_APAImage} showing a fully assembled device. In total, each APA contains 2560 readout wires, with wire spacings of 4.669~mm on the induction planes (denoted by U and V, each consisting of 800 wires) and 4.790~mm on the collection plane (denoted by X, with 960 wires), and plane spacings of 4.75~mm. While this is slightly wider wire spacing than the 3~mm employed by previously described detectors, and studies do show that a smaller spacing (therefore allowing more wires per plane) does result in a marginal improvement ($\approx$ 1~\%) in detector performance, it has been determined \cite{DUNE_IDRv2} that the cost of adding additional wires is not justified for such a small gain. The induction planes have wires orientated at $\pm$35.7\textdegree{} to vertical, and the wires of the collection plane are orientated vertically. Similarly to the LArIAT APA, a fourth plane of 960 vertical wires - denoted by G - is positioned inwards from the U induction plane, i.e. closer to the cathode. This plane is not read out, but acts as an electronic shield and improves the U plane signal quality \cite{DUNE_IDRv2}. Outside the collection plane is a fine mesh, which acts as a ground plane that shields the wire planes from outside interference and improves the wire signal quality by uniformly terminating the drift field. As seen on Figure~\ref{fig:DUNE-SP_Detector}, the middle APA is shared between two drift volumes, and therefore is required to be sensitive on both sides. To achieve this, the induction plane wires are wrapped around the APA frame, as depicted on Figure~\ref{fig:DUNE-SP_APASchematic}. This wrapping also necessitates the choice of the orientation angle: at 35.7\textdegree{}, each induction plane wire only crosses each collection plane wire once across both sides of the APA \cite{DUNE_IDRv2}. This therefore removes any ambiguities that might otherwise occur if an angle of 60\textdegree{} is used, as in previously described detectors.

The readout system to be implemented in DUNE-SP follows the same pattern of cold and warm electronics described previously for SBND - which, as previously mentioned, partly acts as a prototype and testing platform for many DUNE-SP designs and systems. Certain elements of the DUNE-SP wire readout - specifically, the front-end electronics with ASICs comprising amplifiers, pulse-shaping, ADCs and multiplexing - are located on the APA frames themselves, within the LAr environment \cite{DUNE_IDRv2}. Each front-end motherboard is designed to gather the signals from 128 wires, with 20 motherboards therefore required per APA. Cold cables will then carry the collected signals out of the cryostat through feedthroughs, with each feedthrough serving 2 APAs (40 motherboards). The wire biases are also provided through these feedthroughs, as are the connections to and from the PDS (described below) \cite{DUNE_IDRv2}. Additional signal processing will be performed in WIECs mounted on the external sides of the feedthroughs, before the readouts are passed to the off-detector acquisition systems. 

Due to the layout of the DUNE-SP drift volumes, the PDS is extremely constrained in terms of design and available hardware. Since any light-sensitive detectors must of course have an unobstructed view of the drift volumes but be located outside the volumes themselves, the PDS components must be situated within the APA frames, which are only 12~cm across \cite{DUNE_IDRv2}. This therefore rules out the use of large-diameter PMTs. At the time of writing, the final design of the DUNE-SP PDS is still to be confirmed, but two families of light collection devices are under consideration due to their size and relatively high photon detection efficiencies: (X-)ARAPUCA (\cite{ARAPUCA} and \cite{XARAPUCA}, which are to be used on SBND as previously noted) and optical lightguides (similar in concept to those in use in the MicroBooNE detector). Both of these devices will, by their very design, naturally incorporate TPB wavelength shifter, but additional visible light yield may be found (if required) by installing a TPB-coated reflective foil on the CPAs \cite{DUNE_IDRv2}, much like in the LArIAT and SBND detectors. The light collection device that is eventually implemented in DUNE-SP must of course be coupled to a photo-sensitive detector of some nature, and studies are also underway to find and procure a device (or devices) that satisfies the DUNE-SP physics, design and operational requirements \cite{DUNE_IDRv2}.

The colossal scale of the DUNE far detectors - both individually and when considered as a single experiment - will allow DUNE to study a broad range of Particle Physics topics at unparalleled high precision and with high statistics \cite{DUNE_IDRv1}. The primary focus will be the study of neutrino oscillations, including measurement of the $\theta_{23}$ mixing angle, the neutrino mass hierarchy, and testing for the existence of sterile neutrinos. Other topics of interest include (but are by no means limited to) supernovae neutrinos, interaction cross-sections and dark matter searches. Two of the far detector modules are tentatively expected to begin operating in the late 2020s, with the remaining two modules following later as development and funding allow. The order in which the single- and dual-phase modules begin operation is not set, but will depend on the development, operation and results from ProtoDUNE-SP and ProtoDUNE-DP.

\subsection{ProtoDUNE-SP}
\label{subsec:ProtoDUNE_SP}
With an active TPC volume measuring 6 $\times$ 7 $\times$ 7.2~m and containing 411~tonnes of LAr \cite{ProtoDUNE-SP_TDR}, the ProtoDUNE-SP detector is only a fraction of the size of one DUNE far detector module, but nevertheless it is currently the largest single-volume LArTPC ever operated (remembering that the 476~tonnes of LAr at ICARUS T600 are split across two separate volumes). Since late 2018, ProtoDUNE-SP has been in operation at the H4-VLE (``Very Low Energy'') beamline located in the CERN North Area, receiving $e^{-}$, $\mu^{\pm}$, $\pi^{\pm}$ and $p$ at momenta ranging from 0.5 to 7~GeV/c \cite{ProtoDUNE-SP_FirstRes}.

\begin{figure}[b!]
\vspace{2mm}
\centering
\includegraphics[width=0.70\textwidth]{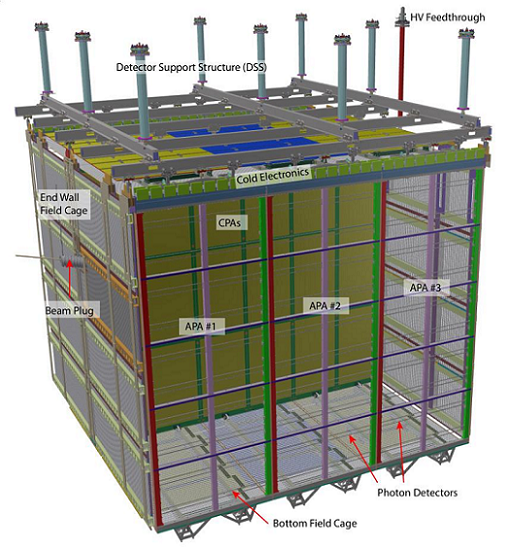}
\caption{The ProtoDUNE-SP LArTPC, with key components labelled. Taken from \cite{ProtoDUNE-SP_TDR}.}
\label{fig:ProtoDUNE-SP_Detector}
\end{figure}

Figure~\ref{fig:ProtoDUNE-SP_Detector} shows the layout of the ProtoDUNE-SP TPC. The detector effectively operates as half of DUNE-SP - with two, rather than four, back-to-back drift volumes of 3.6~m width sharing a common set of CPA sections. As discussed above, the ProtoDUNE-SP CPA sections (previously shown in Figure~\ref{fig:DUNE-SP_FieldCage} (left)) are identical in design and manufacturing to those to be used in DUNE-SP, but ProtoDUNE-SP uses three such sections placed side by side - for a total of 18 1.2 $\times$ 2 m Kapton-laminated FR4 RP elements arranged in a 6 $\times$ 3 grid \cite{ProtoDUNE-SP_TDR}. The drift field is provided via the same field cage assemblies depicted in Figure~\ref{fig:DUNE-SP_FieldCage} (right), and both the nominal cathode bias (-180~kV) and the drift field (500~V/cm) match the design values for DUNE-SP \cite{ProtoDUNE-SP_TDR}. Each of ProtoDUNE-SP's ``APA faces'' consists of three of the 6 $\times$ 2.3 m APAs previously shown in Figure~\ref{fig:DUNE-SP_APAImage} arranged side by side (labelled as ``APA \#1'', ``APA \#2'' and ``APA \#3'' for the detector face shown in Figure~\ref{fig:ProtoDUNE-SP_Detector}), with the four wire planes biased at -665 (shielding plane G), -370 (induction plane U), 0 (induction plane V) and +820~V (collection plane X). It is of course clear that the double-sided sensitivity of the DUNE-SP APAs is not required in ProtoDUNE-SP, but the complete assemblies are nonetheless used so as to validate the design under full operating conditions, and to save on the cost and time of manufacturing dedicated single-sided APAs for ProtoDUNE-SP. The presence of the unused side of wire planes also in fact aids the detector, by collecting and blocking any ionisation occurring outside the TPC from entering \cite{ProtoDUNE-SP_TDR}.

\begin{figure}[b!]
\vspace{2mm}
\centering
\includegraphics[width=0.60\textwidth]{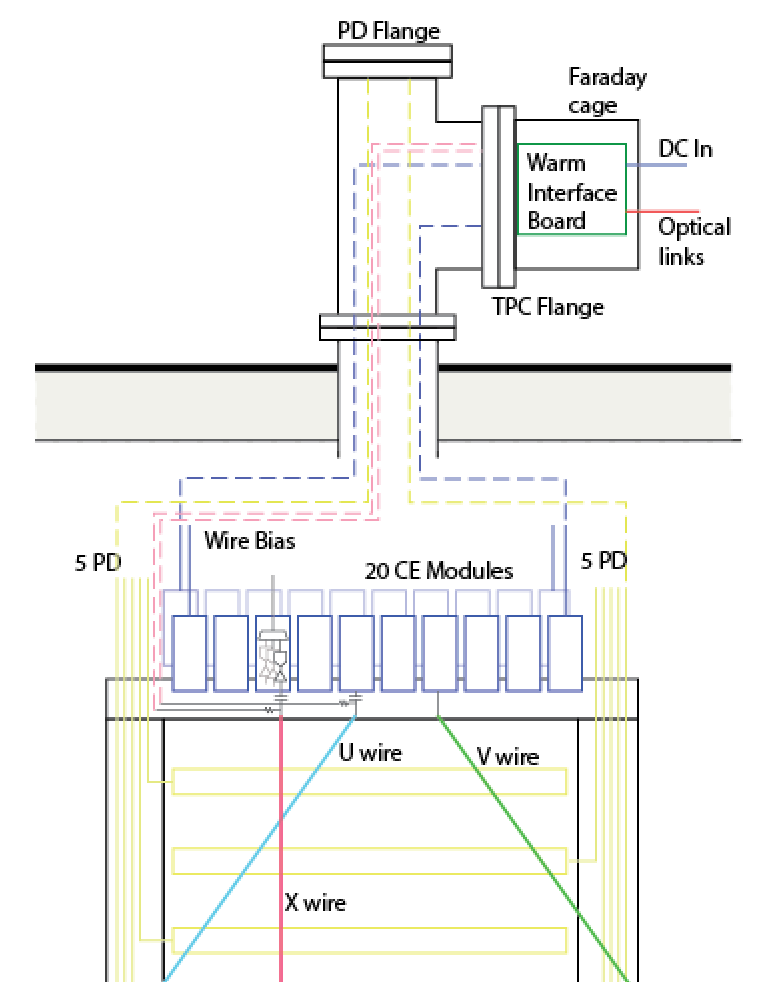}
\caption{A schematic view of the APA readout electronics in ProtoDUNE-SP. Front-end motherboards, 20 of which are mounted on each APA, collect the wire signals and pass them to the WIECs mounted on the external sides of the cryostat feedthroughs. Wire biases and connections to and from the PDS are also provided through the feedthroughs. An identical scheme is anticipated to be used by DUNE-SP. Adapted from \cite{ProtoDUNE-SP_TDR}.}
\label{fig:ProtoDUNE-SP_Readout}
\end{figure}

Signals from the combined 15,360 readout wires are carried out of the TPC via the same system of electronics envisioned for DUNE-SP: a series of front-end motherboards mounted on the APAs gather signals from 128 wires each (with 20 motherboards per APA), and cryostat feedthroughs act as the interface between the cold and warm electronics \cite{ProtoDUNE-SP_TDR}. This is illustrated schematically in Figure~\ref{fig:ProtoDUNE-SP_Readout}. The entire electronics system has undergone a detailed and systematic testing and commissioning program during installation and subsequent operation of the ProtoDUNE-SP detector \cite{ProtoDUNE-SP_ElecPerf}, which showed a 99.7\% working channel efficiency and higher-than-expected operating performance, with low noise and high signal-to-noise ratios on all three wire planes. This bodes extremely well for the eventual implementation of the much larger DUNE-SP APA readout systems using the same components and methods. The detector as a whole has proved to be more than capable of operating at nominal conditions over long periods, with no instabilities or breakages reported on the cathode or field cage elements, and only a very small number of channels failing during the approximately 18 months of data-taking \cite{ProtoDUNE-SP_PerfRes}.

\begin{figure}[h]
\vspace{2mm}
\centering
\includegraphics[width=0.60\textwidth]{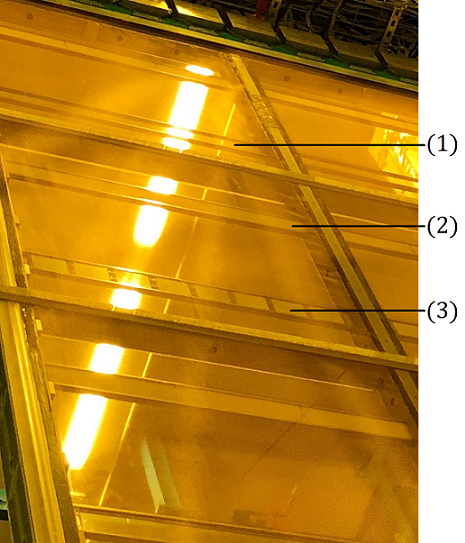}
\caption{The three sub-types of photon detection modules installed behind the wire planes of a ProtoDUNE-SP APA: 1) dip-coated light guides, 2) double-shift light guides, and 3) ARAPUCAs. The ``top'' of the APA (where the front-end electronics are situated) is at the top of the image. Adapted from \cite{ProtoDUNE-SP_PerfRes}.}
\label{fig:ProtoDUNE-SP_PDS}
\end{figure}

As mentioned previously, the choice of hardware to be used in the DUNE-SP PDS has yet to be made, but the ProtoDUNE-SP PDS plays a key part in this by allowing the various devices under consideration to be tested, used and characterised under nominal operating conditions. Each of the six ProtoDUNE-SP APAs can hold ten PDS ``modules'' behind the wire planes, allowing for a total of 60 modules to be installed \cite{ProtoDUNE-SP_PerfRes}. The majority (58) of these are acrylic lightguides, with half of them featuring a ``dip-coated'' design (where each light guide is dipped in a liquid coating containing TPB) and the other half using a ``double-shift'' approach (in which the acrylic is both covered in TPB-impregnated plates and also doped with a secondary wavelength shifter). The two remaining PDS modules are based on the ARAPUCA technology. An example of the module positioning in one APA is shown in Figure~\ref{fig:ProtoDUNE-SP_PDS}. The devices are all read out using silicon photosensors, with three models in use across the 60 PDS modules \cite{ProtoDUNE-SP_PerfRes}, and the signals from these sensors are carried out of the cryostat via the signal feedthroughs, as shown previously in Figure~\ref{fig:ProtoDUNE-SP_Readout}. The PDS has undergone extensive commissioning, calibration and operation (details of which can be found in \cite{ProtoDUNE-SP_PerfRes}), but it has been found that the ARAPUCA modules out-performed both types of lightguides as well as the required specifications for the DUNE-SP PDS - showing excellent collection efficiency, a high photon yield (that is, the number of detected photons per MeV of deposited energy), time resolution on the order of a few tens of nanoseconds, and linear behaviour across a wide range of incident particle energies.

The entire detector assembly described above is housed in a 11.404 $\times$ 11.404 $\times$ 10.756~m membrane cryostat \cite{ProtoDUNE-SP_TDR}. This type of cryostat - also used at SBND, as previously noted - has a long history of use in the transport and storage of liquefied natural gas, and is particularly suited for extremely large cryogenic vessels. The ProtoDUNE-SP cryostat consists of a steel warm outer structure enclosing two membranes (one made of stainless steel and the other of an aluminium composite) and multiple insulation layers. Dedicated penetrations through the cryostat walls are used for the beam plug, which allows incoming particles to travel unimpeded into the active LAr volume, and the LAr extraction pipe that forms part of the purification system (described below). The detector itself is mounted onto a ``detector support structure'' (previously shown in Figure~\ref{fig:ProtoDUNE-SP_Detector}), which is attached to and mechanically supported by the outer cryostat structure \cite{ProtoDUNE-SP_TDR}. A system of cosmic ray taggers is positioned on the upstream and downstream faces of the cryostat (through which the beam enters and exits the detector respectively). These are built from plastic scintillator, and allow the tagging of not only cosmic muons, but also the ``halo'' muons originating from the beam's production target. Due to ProtoDUNE-SP's location at surface level, there is a relatively large cosmic muon flux, and this large population of events has been used as part of the detector calibration studies \cite{ProtoDUNE-SP_TDR}.

The cryogenics systems used by ProtoDUNE-SP \cite{ProtoDUNE-SP_TDR} have been developed in concert with efforts for ProtoDUNE-DP, SBND and ICARUS T600, as they broadly share very similar requirements and specifications. This common approach allows for the pooling of resources and expertise, and will lead to a standardisation of methods and designs, reducing the cost and time that might be otherwise incurred across all four projects. Parts of the ProtoDUNE-SP system - specifically, the commercial provision and onsite storage of liquid argon and nitrogen - are shared with ProtoDUNE-DP due to their close proximity at the CERN North Area. During normal operation of the ProtoDUNE-SP detector, the LAr undergoes continuous purification in a system of external filters containing molecular sieves and copper-coated catalysts, which act to remove water and oxygen contamination respectively \cite{ProtoDUNE-SP_TDR}, with recirculation pumps moving the LAr between the cryostat and filters. Gaseous argon from boil-off within the detector is also recycled back via an external condenser. This system has proved more than capable of achieving sustained high levels of purity, and therefore long electron lifetimes. The maximum drift time in the 3.6~m wide drift volumes is 2.25~ms \cite{ProtoDUNE-SP_PerfRes}, but even during early beam operation, lifetimes of 6~ms had already been achieved \cite{ProtoDUNE-SP_FirstRes}. Within just a few weeks of continuous recirculation and purification, this had improved to consistently exceed 20~ms, achieving a maximum of approximately 89~ms \cite{ProtoDUNE-SP_PerfRes} - corresponding to an impurity level of just 3.4~ppt $O_{2}$ equivalent. (For comparison, the DUNE-SP specification calls for impurity levels of no more than 100~ppt.)

The successful operation of ProtoDUNE-SP at CERN has proved that the underlying design of large single-phase LArTPCs is more than capable of satisfying the requirements for high precision physics analysis, with this experiment matching, and in many areas exceeding, the specifications and results expected from it. However, just as importantly for the purposes of DUNE-SP and future detectors, it has also shown that the methods of manufacturing, assembling, commissioning and operating such detectors are equally up to the task.

\subsection{Pixelated Charge Readout and the ArgonCube Program}
\label{subsec:ArgonCube}
Wire-based anode planes are undoubtedly the most established technology for charge readout in single-phase LArTPCs, with decades of research and refinement behind today's designs and implementations. However, recent developments in reliable and inexpensive cryogenic electronics - such as those used for the front-end components in the SBN and DUNE programs, as previously described - have allowed for the realisation of cryogenic pixelated charge readout. Non-cryogenic implementations of this type of device have been used for almost two decades in gaseous TPCs \cite{PixelCharge_GasTPC}, and typically feature more readout channels than wire-based readouts, as well as digitisation and multiplexing electronics situated very close to the pixels themselves, if not directly bonded to them. The underlying advantage of a pixelated charge readout is that a single anode plane is able to provide full 3D reconstruction of particle tracks in a TPC (with two coordinates provided by the positions of the triggered pixels, and the third being calculated from their relative trigger times), while not being susceptible to the ambiguities that can result from the use of wire-based anode planes. Such ambiguities can occur, for example, in high particle rate environments, where the relatively long readout windows required to fully capture all of the drifting electrons can lead to event pileup, and therefore multiple overlapping signals on the wires. However, usage of a pixelated charge readout at cryogenic temperatures requires not only that the materials and electrical connections be able to mechanically withstand the cold environment, but also that they be able to operate with very low power consumption, in order to minimise heat dissipation.

In recent years, various concepts for pixelated charge readout in LArTPCs have been proposed and demonstrated, with a number of them coming through, or in collaboration with, the ArgonCube program \cite{ArgonCube_Website}. This R\&D program is broadly focused on the design and construction of the LArTPC sub-system of the DUNE near detector, with a primary goal being the implementation of a modular detector design that splits a generic large LArTPC volume into multiple smaller ones, each effectively operating as an independent TPC \cite{ArgonCube_Website, ArgonCube_Presentation}. Such a design - although more complicated in terms of construction and operation - has a number of advantages over a single monolithic detector, such as reduced purity and HV requirements (since the maximum electron drift length per TPC will be smaller), and reduced electron diffusion and scattering. In parallel with this, the development of field cages constructed from thin resistive film is aimed at reducing the amount of uninstrumented, non-active volume between the TPC modules \cite{ArgonCube_Presentation}.

\begin{figure}[b!]
\vspace{2mm}
\centering
\includegraphics[width=0.65\textwidth]{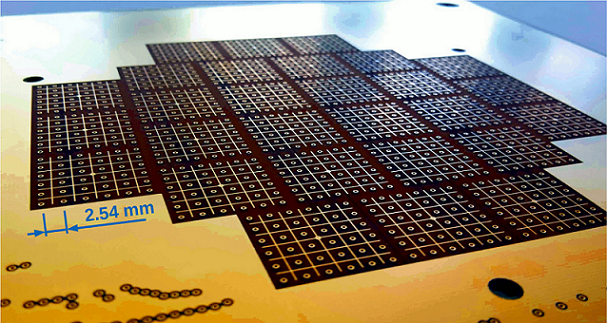}
\caption{The first pixelated charge-capture anode plane developed for the ArgonCube program. It consists of a standard multi-layered PCB, with 1008 900~$\mu$m vias with 2.54~mm spacing acting as pixels. The vias are grouped into 28 6 $\times$ 6~pixel ``regions of interest'' for the purposes of multiplexed readout, and the pixelated area is 100~mm in diameter. Taken from \cite{ArgonCube_FirstDemo}.}
\label{fig:ArgonCube_FirstAnode}
\end{figure}

Initial efforts at developing a pixelated charge readout for LArTPCs began in 2016 \cite{ArgonCube_FirstDemo}. The charge-capture anode plane used in this iteration, shown in Figure~\ref{fig:ArgonCube_FirstAnode}, was a standard multi-layered PCB, with 1008 900~$\mu$m vias (electrical pathways through the PCB layers) acting as the ``pixels''. The pixel-to-pixel spacing was 2.54~mm, and the pixelated area had a diameter of 100~mm, with each pixel directly coupled to a cryogenic preamplifier. The ASICs associated with these preamplifiers were the same as those used in the MicroBooNE and LArIAT detectors - i.e. they were designed for wire-based readout, and so did not have integrated digitising or multiplexing capabilities. This therefore necessitated the use of external digitisers, as well as a bespoke multiplexing algorithm that allowed the number of readout channels to be reduced from 1008 (one per pixel) to just 64 \cite{ArgonCube_FirstDemo}. The LArTPC used for this demonstration was located at the University of Bern, and featured a cylindrical active volume measuring 10~cm in diameter and 60~cm in height (along the drift direction), operating with a drift field of 1~kV/cm \cite{ArgonCube_FirstDemo}. Analysis of cosmic muon events captured by the detector showed that a pixelated charge-capture anode plane is indeed a viable method of reading out a LArTPC, with good a signal-to-noise ratio and precise 3D reconstruction of particle tracks. However, certain limitations were observed, largely related to the aforementioned use of wire-based readout electronics \cite{ArgonCube_FirstDemo}. This made it clear that any further research into pixelated charge readouts would have to include development of dedicated electronics.

\begin{figure}[h]
\vspace{2mm}
\centering
\includegraphics[width=0.50\textwidth]{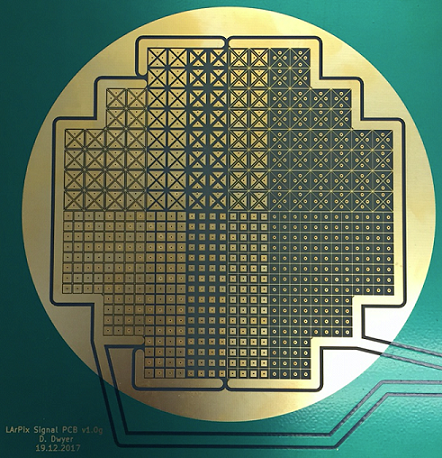}
\caption{The 832-pixel anode plane developed by LArPix. A variety of pixel shapes and sizes were used, in order to evaluate the optimal pixel characteristics. The overall area covered by the pixels is 10~cm in diameter. A second plane, with 512 pixels covering a 4.8 $\times$ 9.6~cm area was also produced and operated. Taken from \cite{ArgonCube_LArPix}.}
\label{fig:ArgonCube_LArPix_Anode}
\end{figure}

Bespoke per-pixel electronics were introduced by the LArPix readout system in 2018 \cite{ArgonCube_LArPix}. In contrast to the vias used by the first iteration, the pixels used by LArPix were gold-plated copper pads with 3~mm spacing, etched onto the surface of a standard dual-layer ``pixel PCB''. Two pixel PCBs were used during the development and operation of LArPix - one containing 832 pixels and an active area 10~cm in diameter (shown in Figure~\ref{fig:ArgonCube_LArPix_Anode}), and the other with 512 pixels covering a 4.8 $\times$ 9.6~cm area. Both used pixels of varying size and shape, for the purposes of evaluating the optimal pixel characteristics. The connected ASICs were located on a separate ``data PCB'' above the pixel PCB, with a shielding layer in between to reduce cross-talk between the PCBs, and with 32 pixels wire-bonded to each ASIC \cite{ArgonCube_LArPix}. The ASICs and PCBs were all manufactured using standard commercial fabrication methods. Central to each pixel's individual readout was a charge-sensitive amplifier connected to a discriminator and digitiser. As the charge built on a pixel, the amplifier's output signal would also increase, and if it subsequently passed a threshold set by the discriminator, it would be passed to the digitiser, after which the amplifier would be reset. This system is therefore the basis of a ``self-triggering'' pixel readout. (A full description of the LArPix ASICs and associated electronics is given in \cite{ArgonCube_LArPix}.) A multiplexing scheme that allowed all of the ASICs to be controlled and read out using a single input and output line respectively was also designed and evaluated.

\begin{figure}[ht]
\centering
\includegraphics[width=0.45\textwidth]{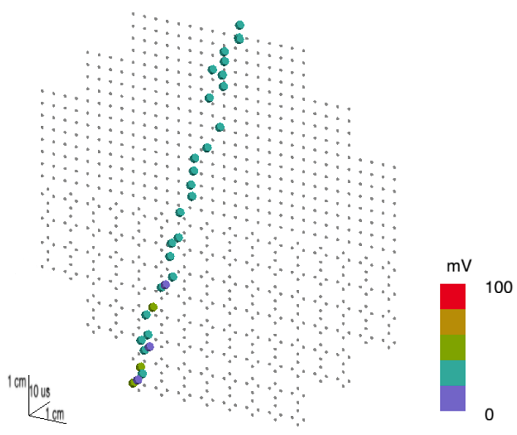}
\hspace{1mm}
\includegraphics[width=0.45\textwidth]{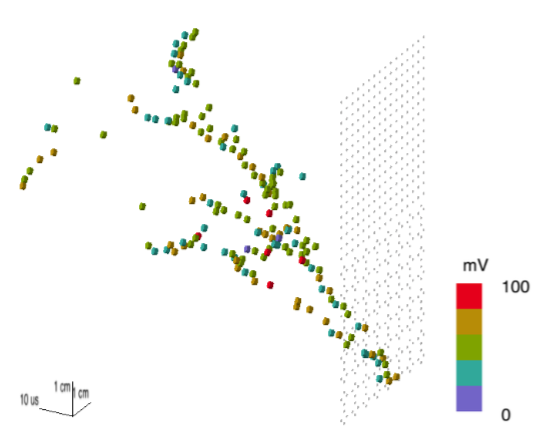}
\caption{(Left) A cosmic muon track as seen by the 832-pixel LArPix readout. (Right) An electromagnetic shower as seen by the 512-pixel readout. In both cases, the positions of the dots on the planes correspond to the pixels that were hit, and the colours of the dots represent the (estimated) charge incident on the pixels. The distance of the dots from the pixel planes is inferred from the relative pixel trigger times. These images represent the raw pixel outputs - no noise filtering or amplification has been performed. Both adapted from \cite{ArgonCube_LArPix}.}
\label{fig:ArgonCube_LArPix_Events}
\end{figure}

The LArPix pixelated charge readouts were tested on two detectors, both at the University of Bern: the 832-pixel iteration was used on a 10~cm diameter, 10~cm tall LArTPC, and the 512-pixel version on the same 10~cm diameter, 60~cm tall setup previously used by the 2016 iteration. Analysis of the collected cosmic muon events from both versions demonstrated the excellent capability of the readouts, showing clean and unambiguous identification of particle tracks with no requirement for noise filtering or external signal amplification, as seen in Figure~\ref{fig:ArgonCube_LArPix_Events}. Testing also demonstrated that the readout electronics were stable and reliable over periods of up to a week of continuous operation, and exhibited a very low noise rate ($<$ 500 electron-equivalent total charge across the entire system) and power consumption ($<$ 100 $\mu$W per pixel) \cite{ArgonCube_LArPix}. Some performance limitations were identified - such as the impact of signal shaping on the precise time of threshold crossing in the discriminator, and the loss of integrated charge during the dead-time between threshold crossing and the amplifier resetting - and further development of the ASIC and associated readout electronics is aimed at addressing such issues.

\begin{figure}[b!]
\vspace{2mm}
\centering
\includegraphics[height=0.35\textheight]{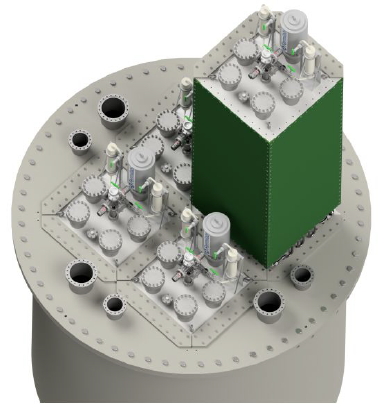}
\caption{A 3D model of the ArgonCube 2 $\times$ 2 demonstrator, showing the positions of the four TPC modules (one partially removed). The demonstrator has an overall active volume of 1.2 $\times$ 1.3 $\times$ 1.3~m \cite{ArgonCube_Presentation2}. Taken from \cite{ArgonCube_Presentation}.}
\label{fig:ArgonCube_Detector}
\end{figure}

\begin{figure}[h!]
\vspace{2mm}
\centering
\includegraphics[width=0.85\textwidth]{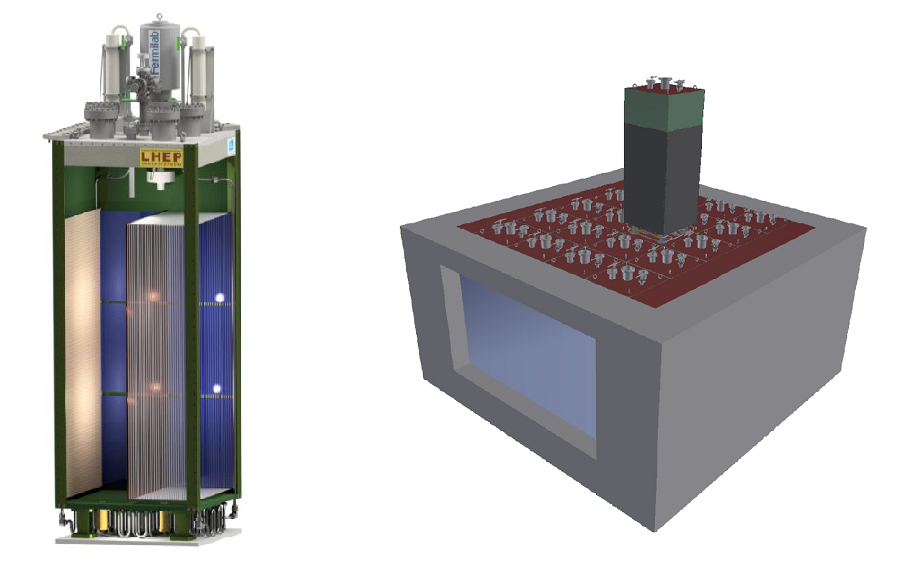}
\caption{(Left) A 3D model of one module of the proposed LArTPC sub-system of the DUNE near detector. The basic design is that of two 0.5~m wide horizontal single-phase drift volumes arranged back-to-back, with a shared cathode orientated vertically in the middle and pixelated charge readout anode planes on the corresponding outer faces. Each module will possess its own dedicated LAr recirculation, HV and photon detection systems.  (Right) The envisioned design of the LArTPC sub-system of the DUNE near detector, consisting of 20 modules arranged in a 4 $\times$ 5 grid encompassing an active volume measuring 3 $\times$ 4 $\times$ 5~m. Both taken from \cite{ArgonCube_Presentation2}.}
\label{fig:ArgonCube_DUNE-ND_Schematic}
\end{figure}

Development and implementation of the ArgonCube modular detector design continues alongside the LArPix program. A demonstrator, which uses a 2 $\times$ 2 array of TPCs and is shown in Figure~\ref{fig:ArgonCube_Detector}, is in the process of being assembled, with the cryostat and first module having been completed and commissioned in 2019 \cite{ArgonCube_Presentation}. Once complete, this detector will be operated on the NuMI beamline at Fermilab in order to fully characterise its functionality and operation in a similar environment to that which will be experienced by the DUNE near detector. Design and physics studies are also underway for the eventual extension of the modular design to the scale of the DUNE near detector \cite{ArgonCube_Presentation}, as depicted in Figure~\ref{fig:ArgonCube_DUNE-ND_Schematic}, and proposals have also been made for a kiloton-scale iteration, which is envisioned as one of the DUNE far detector modules \cite{ArgonCube_Kilotonne}.

\begin{figure}[ht]
\centering
\includegraphics[width=0.70\textwidth]{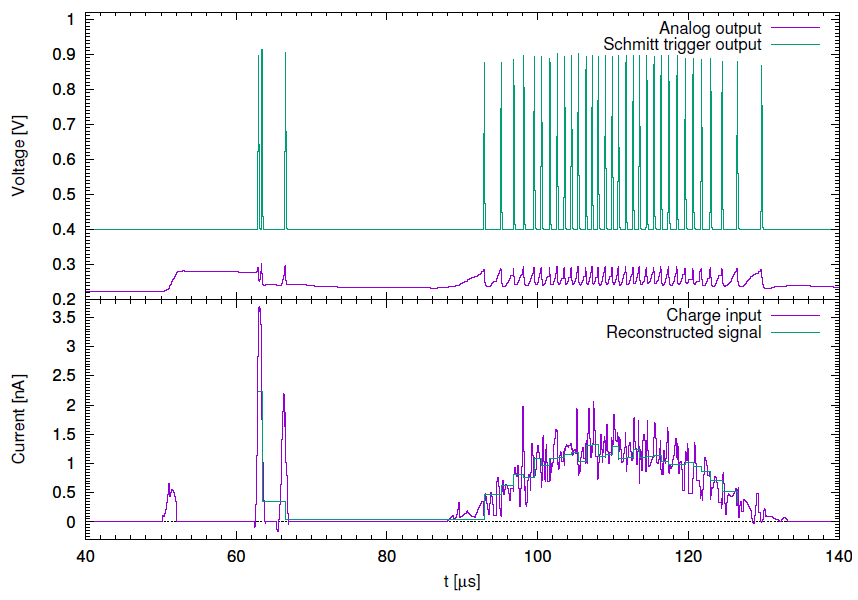}
\caption{A simulation of the Q-Pix response and corresponding waveforms for an integrated charge threshold of 0.3~fC. (Top panel) The integration of charge on the amplifier (purple) and the corresponding Schmitt trigger reset signals (green). (Bottom panel) The simulated input current (purple), and the reconstructed waveform (green). The shape of this waveform is directly determined from the distribution of charge integration on the amplifier, since the accumulated charge between resets is known and fixed. Taken from \cite{ArgonCube_QPixConcept}.}
\label{fig:ArgonCube_QPix_Waveform}
\end{figure}

An alternative approach to signal digitisation in pixelated charge readouts has been recently proposed by the Q-Pix consortium \cite{ArgonCube_QPixConcept}. As with LArPix, each pixel in the Q-Pix design is connected to an ASIC containing a self-triggering circuit: the charge collected on the pixel is integrated on a charge-sensitive amplifier, the integral is compared to a pre-set threshold, and once this threshold is exceeded the amplifier is reset (in this design, via a Schmitt trigger). However, in Q-Pix a local clock is also present in the ASIC, and when the threshold is exceeded, the timestamp on this clock is read out and stored. As long as charge continues to build on a pixel, the circuit will continue to output timestamps, and the time difference between successive timestamps (referred to as the ``reset time difference'' (RTD)) therefore represents the time required for the pixel to collect a fixed and known amount of charge - the ``time-to-charge'' \cite{ArgonCube_QPixConcept}. The pattern of RTDs across many pixels on a charge-capturing anode plane can then be related to the spatial and temporal shape of the incident charge, and therefore what type of event produced it. For example, a short series of small RTDs produced by many pixels indicates that a lot of charge was incident in a short space of time (i.e. the pixels' amplifiers were reset multiple times in quick succession), corresponding to a spatially wide but temporally localised ionisation of the LAr, i.e. a particle track. On the other hand, background events or noise - which generally produce low-rate, highly localised ionisation - will be observed as large RTDs in only a few pixels across the plane. A conventional waveform, equivalent to that produced by the digitisers in LArPix, can also still be produced in Q-Pix, as illustrated in Figure~\ref{fig:ArgonCube_QPix_Waveform}.

Due to the vast majority of the pixels' electronics being effectively inactive at any given moment, the power consumption and data rates of a Q-Pix based anode plane would both be very low, while the obvious difference between the responses to signal and background (and noise) leads to a very high signal-to-noise ratio \cite{ArgonCube_QPixConcept}. Initial designs and simulation of the macroscopic Q-Pix readout - that is, the network of ASICs that would be required to read out an entire plane of pixels - have produced promising results, including dynamic communication between the ASICs (required for correlated timestamp readout) and robust fault protection. Work is ongoing, both within the Q-Pix consortium and also in collaboration with LArPix, to further refine the Q-Pix concept \cite{ArgonCube_QPixPresentation}.

\subsection{Vertical Drift}
\begin{figure}[ht]
\centering
\includegraphics[width=0.95\textwidth]{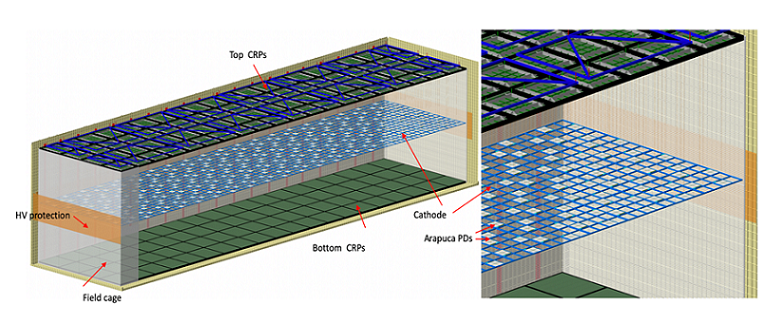}
\caption{A schematic of the Vertical Drift LArTPC design, as envisioned for a DUNE far detector module. Two 6.5~m tall drift volumes are orientated vertically, with a common cathode in between, and charge readout planes (CRPs, in the form of 160 perforated PCB anodes) are positioned above and below the top and bottom drift volumes respectively. ARAPUCA photon detectors will be embedded within the cathode for direct observation of scintillation light. Taken with permission from \cite{VertDrift_Nessi}.}
\label{fig:VertDrift_Detector}
\end{figure}

Although existing (Proto)DUNE-SP technologies, manufacturing and assembly procedures have proved to be extremely successful and will serve the neutrino sector well for many years to come, new ideas for the design of large-scale LArTPCs continue to be proposed and developed. One very recent avenue of research is known as ``Vertical Drift'' \cite{VertDrift_Nessi} - hereafter referred to as ``VD'' for convenience, and shown in Figure~\ref{fig:VertDrift_Detector} as envisioned for a DUNE far detector module. Taking a number of cues from the design of dual-phase detectors (discussed in more detail in Section~\ref{sec:dualPhase}), the VD concept represents a shift from the established layout of single-phase LArTPCs, utilising vertical rather than horizontal electron drift. However, experience with the design and construction of ProtoDUNE-DP has shown that a vertically orientated layout is generally more efficient in terms of active LAr volume usage, due to having only one or two drift volumes, and therefore fewer components and smaller uninstrumented regions in comparison to the multiple volumes present in the larger horizontally orientated single-phase detectors. This makes such vertically orientated experiments easier to assemble \cite{VertDrift_Nessi}, thus reducing required costs and time. Further reductions can also be achieved by making use of the many years of research and development that have already been performed for the (Proto)DUNE experiments - i.e. by reusing common designs, components and technologies that have already been proven and characterised.

A DUNE far detector utilising the VD design will consist of two drift volumes positioned one on top of the other, with a combined active LAr mass of 14,740~tonnes and a common horizontal cathode between them. This cathode is envisioned to consist of a 58~mm thick FR4 frame with welded wire meshes covering the top and bottom surfaces \cite{VertDrift_Nessi}. The cathode bias will be provided through a HV feedthrough and extension system similar to that developed for (Proto)DUNE-DP (see Section~\ref{subsec:ProtoDUNE_DP}), but due to the vertical drift length being 6.5~m, the required bias in the VD design will be 300 to 350~kV - almost twice that of the (Proto)DUNE-SP detectors, but approximately equal to that of ProtoDUNE-DP. Similarly, the field cage elements in the VD design are also envisioned to be the same as those on SBND, (Proto)DUNE-SP and (Proto)DUNE-DP.

\begin{figure}[ht]
\centering
\includegraphics[height=0.20\textheight]{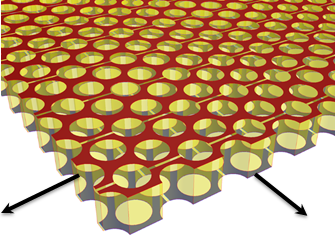}
\hspace{1mm}
\includegraphics[height=0.20\textheight]{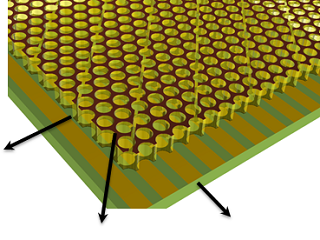}
\caption{(Left) A 2-layer perforated PCB anode, with the layers cut into strips orientated at 90\textdegree{} to each other and perforated by 2.5~mm diameter holes. One layer acts as an induction plane, and the other as a collection plane. (Right) A 3-layer perforated PCB anode, in which the additional layer consists of strips orientated at 45\textdegree{} to the other two, located on a separate PCB and acting as another induction plane. In both images, the arrows denote the direction of strip readout on each layer. Both taken with permission from \cite{VertDrift_Pietropaolo}.}
\label{fig:VertDrift_Anode}
\end{figure}

Recent advances in the design and manufacturing of large-area THGEMs (``Thick Gaseous Electron Multipliers'' - a type of micropattern detector that is widely used as the anode in dual-phase TPCs, and discussed in more detail in Section~\ref{sec:dualPhase}) have allowed for the development of a new type of anode plane that has been proposed for use on a future VD detector: a perforated PCB anode \cite{VertDrift_Pietropaolo}. This type of device, shown in Figure~\ref{fig:VertDrift_Anode} (left), is constructed from a two-layer PCB of 3.2~mm thickness and perforated with 2.5~mm diameter holes, with each layer cut into strips of 5.2~mm width along either the $x$ or $y$ axis. The layers are electrically biased such that one acts as an induction plane, and the other as a collection plane. The PCB then operates in a completely analogous way to a 2-plane wire-based APA: electrons drifting towards it are focused into the holes, inducing bipolar signals on the nearby induction plane strips, and are then captured by the collection plane, creating a unipolar signal on a particular collection strip. An alternate design, shown in Figure~\ref{fig:VertDrift_Anode} (right) includes a second PCB, positioned 10~mm from the first one and consisting of a single layer with 8.7~mm wide strips orientated at 45\textdegree{} to those on the first PCB. With appropriate biasing, this additional PCB acts as a second induction layer, brings the design to equivalence with the 3-plane wire-based APAs described previously \cite{VertDrift_Pietropaolo}.

\begin{figure}[h]
\centering
\includegraphics[width=\textwidth]{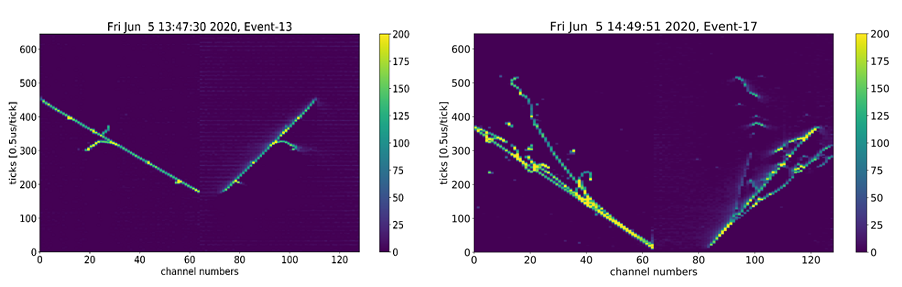}
\caption{The readout strip signals of the 32 $\times$ 32~cm 2-layer perforated PCB anode operating in the CERN 50l LArTPC, for a cosmic muon (left) and an EM shower (right). Each ``pixel'' in these images corresponds to the signal size (ADC units, colour axis) on a given strip (channel number, $x$ axis) at a given time (clock ticks, $y$ axis). The bottom of each image (time = 0) corresponds to the anode position, with the cathode at the top. In both images, the induction plane strips are on the right (channel numbers $\geq$ 63) and the collection plane strips are on the left (channel numbers $<$ 63). Both taken with permission from \cite{VertDrift_Pietropaolo}.}
\label{fig:VertDrift_Events}
\end{figure}

A perforated PCB anode has several advantages over wire-based APAs \cite{VertDrift_Pietropaolo}. Large area PCBs can be produced commercially and in bulk, helping to reduce the costs, manpower and time associated with manufacturing and assembly. This is particularly important to consider for a DUNE far detector implementation, which is envisioned to incorporate a total of 160 3 $\times$ 3.4~m perforated PCB anode assemblies across the two ``charge readout planes'' (CRPs) \cite{VertDrift_Nessi, VertDrift_Pietropaolo}, as shown in Figure~\ref{fig:VertDrift_Detector}. The perforated PCB anode design itself is more robust than tensioned wires, and it may also be possible to integrate the front-end readout electronics within the top PCB layer - in a similar way to how the previously described pixelated charge readout electronics are bonded directly to the pixels that they service. The remainder of the readout components are expected to parallel the systems which have been successfully operated by the ProtoDUNE experiments \cite{VertDrift_Nessi}. A proof-of-principle test of a 2-layer perforated PCB anode has recently been performed using cosmic muon data taken on a 50l LArTPC at CERN \cite{VertDrift_Pietropaolo}. The anode used was 32 $\times$ 32~cm in area, but otherwise identical to the proposed VD design, as were the readout electronics. The readout strip signals, examples of which are depicted in Figure~\ref{fig:VertDrift_Events}, show clear and precise particle tracks, with high signal-to-noise ratio for both MIP and shower-like topologies. A similar test of the 3-layer iteration, also on the 50l detector, is due to take place in early 2021 \cite{VertDrift_Resnati}.

Taking full advantage of the various characterisation studies of different light collection devices already performed for the (Proto)DUNE experiments, the VD concept is envisioned to use a system of ARAPUCAs and associated SiPMs as its PDS \cite{VertDrift_Pietropaolo}. However, unlike previous single-phase detectors which locate these assemblies adjacent to the anode planes, the VD design calls for them to be embedded within the cathode, within the previously noted 58~mm gap between the meshes \cite{VertDrift_Pietropaolo}. This location does require the development (currently ongoing) of new ``power over (optical) fibre'' (PoF) readout and operation electronics for the SiPMs, since there will be insufficient access or space for standard electronics, but the overall PDS design is anticipated to perform just as well as one in which the ARAPUCAs are mounted on the anodes, while using fewer individual devices \cite{VertDrift_Pietropaolo}. An ``extended'' system, using additional ARAPUCAs positioned on the field cage looking inwards, would allow the VD PDS to have 4$\pi$ coverage, greatly improving the detection of low energy events in the detector, and xenon-doping of the LAr (up to the level of $\approx$ 19~ppm) has also been proven to increase the uniformity and amount of collected light, by increasing the optical scattering length and significantly reducing the effects of N$_2$ contamination \cite{VertDrift_Pietropaolo, VertDrift_Resnati}. Xenon-doped LAr has a peak scintillation emission at 175~nm \cite{VertDrift_Pietropaolo}, rather than the 128~nm of pure LAr, and studies are currently underway to optimise the ARAPUCAs and SiPMs for photon collection and detection at this wavelength.

Refinement and characterisation of the various VD design components is in progress, with the assembly and operation of a full-scale ``VD module'' (consisting of a single 3 $\times$ 3.4~m perforated PCB anode assembly, the complete readout system for such a device, and a corresponding cathode with embedded PDS and associated HV systems) planned for the second half of 2021 \cite{VertDrift_Resnati}.

\section{Dual-Phase Experiments}
\label{sec:dualPhase}
Dual-phase LArTPCs are a relatively recent addition to the neutrino sector, having only been given detailed consideration and development since the early 2010s. However, the underlying dual-phase concept itself is by no means a new one, having been used for many years in dark matter searches \cite{DarkMatterDetectors}. The xenon-based ZEPLIN-II \cite{ZeplinII} and ZEPLIN-III \cite{ZeplinIII} experiments pioneered the design and operation of dual-phase TPCs in the early 2000s, and more recent developments include LUX \cite{LUX}, LUX-ZEPLIN \cite{LUXZEPLIN} (both also using xenon) and the argon-based DarkSide program \cite{DarkSide}. One of the key differences between implementations of dual-phase technology in the neutrino and dark matter sectors is that neutrino experiments require high resolution position reconstruction across a relatively large spatial region, i.e. a particle track, whereas dark matter searches call for high precision calorimetry of more point-like energy depositions.

\begin{figure}[ht]
\centering
\includegraphics[width=0.55\textwidth]{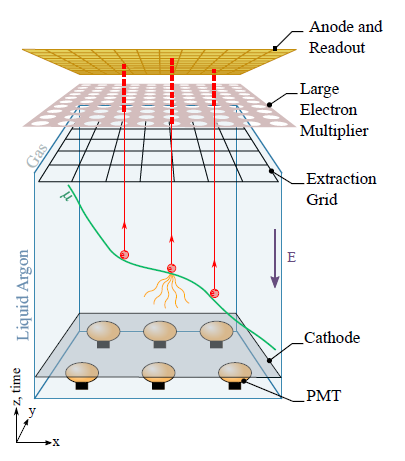}
\caption{The general detection principle of dual-phase LAr detectors with charge readout. This schematic depicts the assembly in the dual-phase DUNE far detector, but the same principle is applicable to any such design. A description of the operation is given in the main text. Taken from \cite{DUNE_IDRv3}.}
\label{fig:DUNE-DP_DetectionPrinciple}
\end{figure}

The general detection principle of dual-phase TPCs with charge readout is shown in Figure~\ref{fig:DUNE-DP_DetectionPrinciple}. Electrons produced during ionisation are drifted upwards towards the surface of the liquid scintillator phase. Just below the surface is positioned an ``extraction grid'', and just above it in the gas phase is a ``(Thick) Gaseous Electron Multiplier'' or ``Large Electron Multiplier'' (abbreviated to (TH)GEM or LEM respectively). (The names ``THGEM'' and ``LEM'' both refer to the same type of device, and when describing the specific hardware used by each experiment henceforth, we will use the name that is predominantly used by that experiment's publications.) THGEMs/LEMs \cite{THGEMReview} are a further development of GEMs \cite{GEMDescription}, and are a type of micropattern detector that typically consists of an insulating layer with conducting material on the top and bottom surfaces that can be individually biased, and many thousands of holes drilled through the entire structure. The extraction grid and bottom THGEM/LEM surface are biased so as to produce an electric field between them (referred to as the ``extraction field'') that is stronger than the drift field, resulting in the electrons being accelerated sufficiently to escape the liquid and enter the gas. The top THGEM/LEM surface is also biased in such a way that the electrons are subject to an ``amplification field'' across the THGEM/LEM, further accelerating them through the holes and inducing electron avalanches. This results in electron multiplication and the production of secondary scintillation light. (The entire process is known as ``Townsend multiplication''.) The proof of concept of a dual-phase argon TPC operating with a GEM and THGEM readout was demonstrated in \cite{DualPhasePoC1} and \cite{DualPhasePoC2} respectively.

In dual-phase detectors which utilise charge readout, the multiplied electrons are finally collected by a wire-based anode plane positioned above the THGEM/LEM. As the extraction grid, THGEM/LEM and anode plane must all be present together in order for charge readout to occur, they are often grouped into a single assembly known as a ``charge readout plane'' (CRP).

\subsection{DUNE (Dual-Phase Far Detector)}
\label{subsec:DUNE_DP}

\begin{figure}[ht]
\centering
\includegraphics[width=\textwidth]{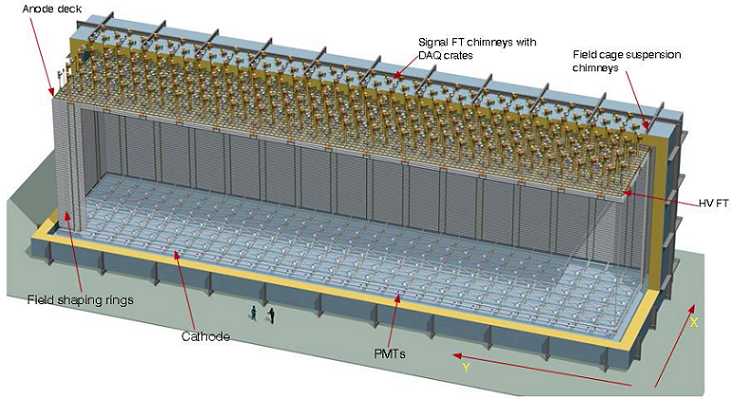}
\caption{A schematic view of the dual-phase DUNE far detector module, with major components labelled. Taken from \cite{DUNE_IDRv3}.}
\label{fig:DUNE-DP_Detector}
\end{figure}

In a similar way to single-phase LArTPCs under the ICARUS program, the development of dual-phase LArTPCs for neutrino detection has followed a path of progressively larger detectors - initially using a 10 $\times$ 10 $\times$ 21~cm TPC volume containing $\approx$ 3~kg of LAr \cite{FirstDualPhase_Part1, FirstDualPhase_Part2}, followed by a 200~kg 40 $\times$ 76 $\times$ 60~cm iteration \cite{SecondDualPhase}. Much of the most recent development has been performed in the context of the dual-phase DUNE program, which includes the 3 $\times$ 1 $\times$ 1~m WA105 and 6 $\times$ 6 $\times$ 6~m dual-phase ProtoDUNE (``ProtoDUNE-DP'') detectors (both described below in Section~\ref{subsec:ProtoDUNE_DP}), as well as the dual-phase DUNE far detector module (``DUNE-DP'') itself. The latter is conceptually depicted in Figure~\ref{fig:DUNE-DP_Detector}.

DUNE-DP will have an active LAr mass of 12,096~tonnes (10,643~tonnes fiducial) \cite{DUNE_IDRv3} contained within a single 60 $\times$ 12 $\times$ 12~m TPC. All of the designs and technologies to be implemented on DUNE-DP will be scaled-up versions of those used by the WA105 and ProtoDUNE-DP experiments, and so the successful construction and operation of these detectors represent a critical step towards the realisation of DUNE-DP.

\subsection{WA105 and ProtoDUNE-DP}
\label{subsec:ProtoDUNE_DP}
\begin{figure}[h!]
\vspace{2mm}
\centering
\includegraphics[width=0.95\textwidth]{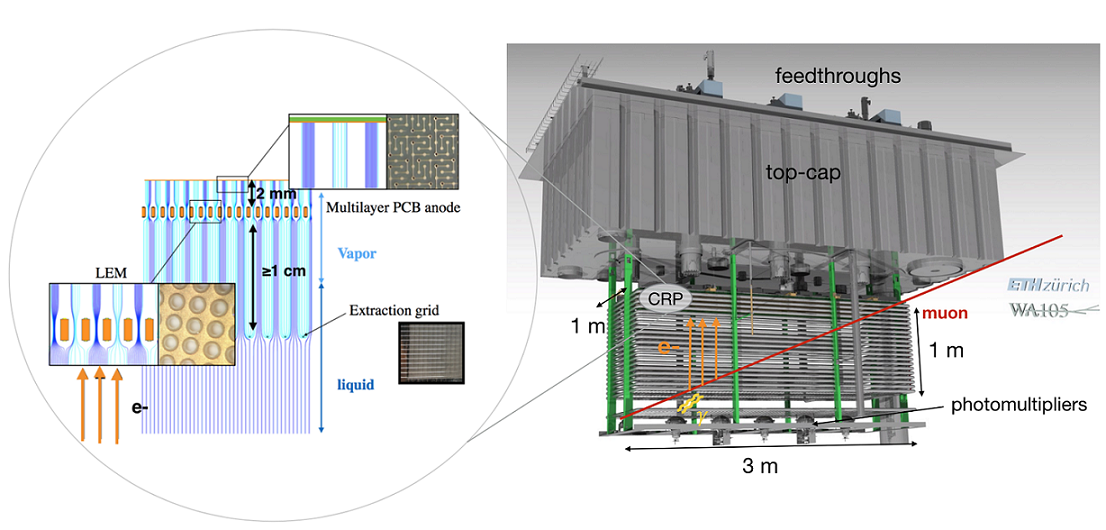}
\caption{A model of the WA105 dual-phase LArTPC, showing the field cage suspended from the 1.2~m thick cryostat top-cap, as well as the various feedthroughs that penetrate through to the interior. The inset shows the structure of the charge readout plane (CRP), described in more detail in the main text. Taken from \cite{WA105_Presentation}.}
\label{fig:WA105_Detector}
\end{figure}

\emph{Authors' Note: at various points during the dual-phase DUNE program, the two prototype detectors have each been referred to by several different names. The smaller of the two, with dimensions of 3 $\times$ 1 $\times$ 1~m, has gone by ``WA105'', ``WA105-DEMO'' and ``WA105-311''. The larger 6 $\times$ 6 $\times$ 6~m iteration has previously also been designated ``WA105'', as well as ``WA105-666'' and ``NP02'', and is now commonly known as ``ProtoDUNE-DP''. For clarity and convenience, in this paper we will use ``WA105'' to exclusively refer to the smaller prototype, and ``ProtoDUNE-DP'' for the larger one.}
\\
\\
Constructed in 2016, the WA105 experiment (so called due to its location at CERN's ``West Area'') is the first of two detectors dedicated to assessing the scalability of the dual-phase LArTPC concept - that is, if a DUNE far detector based on this design could be assembled and operated successfully. To that end, WA105 features a 3 $\times$ 1 $\times$ 1~m TPC \cite{WA105_Presentation} enclosed by a 1~m thick membrane cryostat constructed from layers of glass-reinforced polyurethane foam (for thermal insulation) and plywood (for pressure distribution). The TPC field cage, containing $\approx$ 4.2~tonnes of LAr, is suspended from the 1.2~m thick top-cap of the cryostat, as shown in Figure~\ref{fig:WA105_Detector}.

At the bottom of the WA105 field cage is the cathode - a metallic grid nominally biased at -50~kV via a dedicated HV feedthrough \cite{WA105_Presentation}. This feedthrough has also been successfully operated at -300~kV, in order to test the design for eventual use in the larger ProtoDUNE-DP detector. The 1~m tall field cage consists of 20 field-shaping elements positioned at 5~cm intervals above the cathode, with the drift field being produced by a resistor chain attached to the field-shaping elements. In order to detect the prompt scintillation light for the purposes of event triggering and absolute $z$-positioning, 5 8-inch diameter Hamamatsu R5912-MOD PMTs are positioned below the cathode - three of which are directly coated with TPB, with the remaining two operating in conjunction with TPB-coated acrylic discs \cite{WA105_Presentation2}.

\begin{figure}[h]
\vspace{2mm}
\centering
\includegraphics[width=0.90\textwidth]{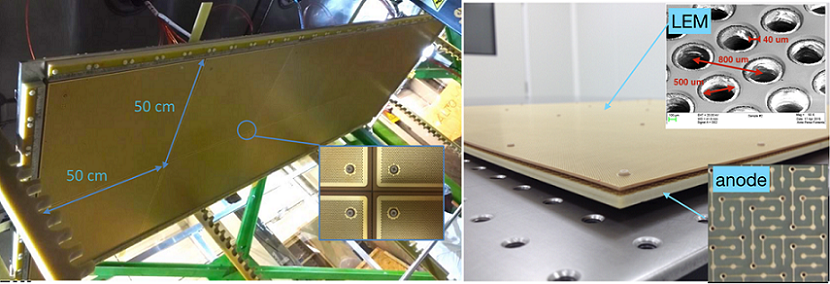}
\caption{(Left) The fully assembled 3 $\times$ 1~m WA105 charge readout plane, viewed from underneath in order to see the bottom surfaces of the individual 50 $\times$ 50~cm LEM units. The extraction grid is not shown. (Right) The ``sandwich'' structure of the combined LEM and anode plane, with a spacing of 2~mm between them. Insets show the fine structure of the LEM (top) and anode (bottom) surfaces. Both taken from \cite{WA105_Presentation}.}
\label{fig:WA105_CRP}
\end{figure}

The 3 $\times$ 1~m WA105 CRP, shown in Figure~\ref{fig:WA105_CRP}, consists of the extraction grid, LEM and anode plane as previously described in Section~\ref{sec:dualPhase}. The extraction grid is a single plane of 100~$\mu$m diameter tensioned wires, with 3.125~mm wire-to-wire spacing in both $x$ and $y$ directions, and held 5~mm below the surface of the LAr \cite{WA105_Presentation}. In contrast, both the LEM (positioned 5~mm above the LAr surface) and anode plane (2~mm above the LEM) are each assembled from 12 50 $\times$ 50~cm ``units'', which are electrically connected to form a single continuous plane for each layer. Each of the LEM units - the fine structure of which is shown in Figure~\ref{fig:WA105_CRP} (top-right inset) - is a commercially manufactured PCB, with holes of 500~$\mu$m diameter (with an additional 40~$\mu$m dielectric rim to reduce the risk of electrical breakdown) and 800~$\mu$m ``pitch'' (centre-to-centre hole spacing). These characteristics have been decided upon following a series of performance tests conducted on a variety of LEM designs \cite{THGEM_Tests}. The extraction and amplification fields are 2 and 30~kV/cm respectively \cite{WA105_Presentation}. Each anode plane unit - which are also commercially produced PCBs - consists of two layers of 3~mm wide strips arranged so as to produce unambiguous 2D ($x$ and $y$) position reconstruction of the electrons incident from the LEM. The pattern of the strips, details of which are shown in Figure~\ref{fig:WA105_CRP} (bottom-right inset), has been optimised to give good position resolution as well as equal charge sharing - and therefore sensitivity - between the $x$ and $y$ views \cite{WA105_Presentation, Anode_Tests}.

Although the field cage is fixed to the underside of the top-cap using eight FR4 supports, the CRP is not in turn directly attached to the field cage. Instead, it is suspended from three cables which pass into the cryostat interior via dedicated feedthroughs \cite{WA105_Presentation}. These cables can be externally adjusted, and so this setup allows the height and relative flatness of the CRP to be optimised independently and without needing to access the rest of the TPC.

Ease of access is also a guiding factor in the design of the WA105 signal feedthroughs. Each one is a 2~m long nitrogen-filled tube sealed at both ends by vacuum-tight PCBs \cite{DUNE_IDRv3, WA105_Presentation}, and rather than mounting the cold front-end electronics boards (which perform initial amplification and multiplexing) directly on the TPC or anode plane, they are located within the feedthrough, in the space between - and connected to - the PCBs. The bottom PCBs are in turn connected to the anode plane readouts, and the top PCBs to the warm intermediate electronics positioned externally. This feedthrough design therefore allows the front-end electronics to be inserted into or removed from the detector without requiring access to the entire interior, while still situating them in thermal contact with the gaseous Ar, where they can benefit from the cold environment \cite{WA105_Presentation}.

\begin{figure}[h]
\centering
\includegraphics[width=0.85\textwidth]{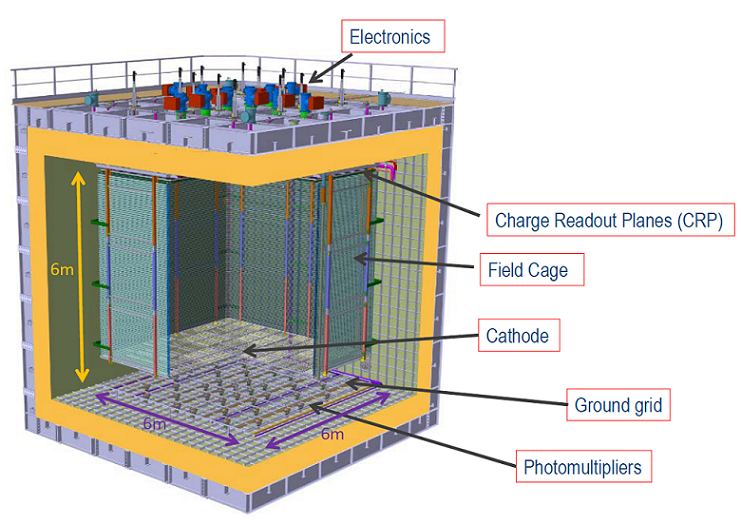}
\caption{A schematic depiction of the ProtoDUNE-DP detector, with major components labelled. Taken from \cite{ProtoDUNE-DP_Presentation}.}
\label{fig:ProtoDUNE-DP_Detector}
\end{figure}

The WA105 detector was operated during 2017, collecting cosmic muon data and performing a detailed characterisation of the LArTPC design, components and operation. The successful completion of this first stage of the dual-phase DUNE R\&D program provided valuable experience and insight for the construction and operation of ProtoDUNE-DP. This detector, shown in Figure~\ref{fig:ProtoDUNE-DP_Detector}, is considerably larger than WA105 - with a drift volume of 6 $\times$ 6 $\times$ 6~m \cite{ProtoDUNE-DP_Presentation} and an active LAr mass of just over 300~tonnes. (In actual fact, this is the size of detector originally proposed as the dual-phase LArTPC demonstrator \cite{ProtoDUNE-DP_original}, before it was appreciated that the construction and operation of a smaller iteration would be beneficial.)

\begin{figure}[ht]
\centering
\includegraphics[width=0.70\textwidth]{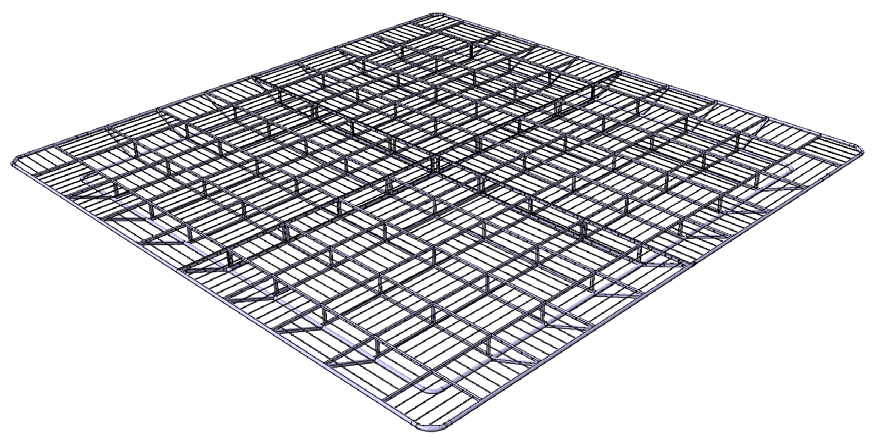}
\caption{A 3D model of the ProtoDUNE-DP cathode plane, which consists of 4 3 $\times$ 3~m units. The grid structure is manufactured from three sizes of stainless steel tubing: the external perimeter is constructed from 60~mm diameter circular cross-section pipes, 20 $\times$ 40~mm oval cross-section pipes make up the internal frame, and fine structure is provided by 12~mm diameter pipes. The dual-phase DUNE far detector's cathode will use an identical grid design, but scaled up to 80 units making a single 12 $\times$ 60~m plane. Taken from \cite{ProtoDUNE-DP_Presentation}.}
\label{fig:ProtoDUNE-DP_Cathode}
\end{figure}

\begin{figure}[h!]
\vspace{5mm}
\centering
\includegraphics[width=0.80\textwidth]{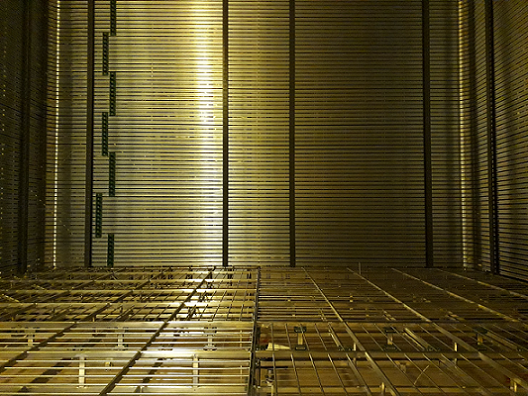}
\caption{A view of the interior of the ProtoDUNE-DP drift volume. The four cathode units can be seen at the bottom, and the field cage (with attached resistor chain elements) is installed on the vertical faces. Taken from \cite{ProtoDUNE-DP_Presentation}.}
\label{fig:ProtoDUNE-DP_TPC}
\end{figure}

Similarly to that of WA105, the ProtoDUNE-DP cathode is a stainless steel grid: the design is shown in Figure~\ref{fig:ProtoDUNE-DP_Cathode}. The entire structure consists of 4 ``units'', each of which measures 3 $\times$ 3~m and are electrically connected (and isolated from each other) via resistors once assembly of the entire cathode is completed. This division of the cathode into smaller units is done in order to reduce the impact of discharges on the surrounding electronics and cryostat structure \cite{DUNE_IDRv3}. The DUNE-DP cathode is envisioned to use an identical design, but with a 12 $\times$ 60~m assembly of 80 units. The layout of the grid results in the cathode having 60\% optical transparency \cite{DUNE_IDRv3}, allowing the placement of PDS components below it. The nominal drift field in ProtoDUNE-DP will be 500~V/cm, provided via a field cage identical in design to those used on SBND and (Proto)DUNE-SP: rolled aluminium profiles mounted onto a modular framework of pultruded fiber-reinforced polymer beams \cite{DUNE_IDRv3}. The interior of the ProtoDUNE-DP drift volume, showing the assembled field cage and cathode, is shown in Figure~\ref{fig:ProtoDUNE-DP_TPC}.

\begin{figure}[h]
\vspace{2mm}
\centering
\includegraphics[width=0.50\textwidth]{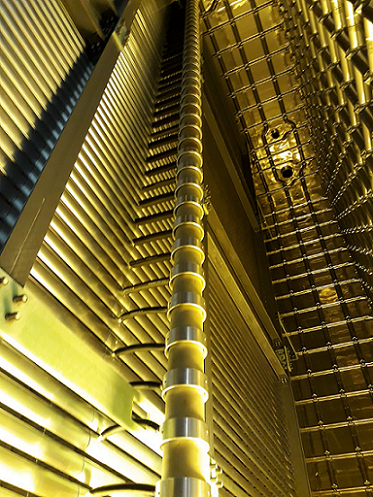}
\caption{The 6~m long ProtoDUNE-DP HV extender, consisting of an inner conducting core that brings the -300~kV bias from the HV feedthrough to the cathode, surrounded by an insulator. Voltage degrader rings are positioned around the insulator and connected to the nearby field cage elements in order to reduce any electric field in this region to zero. The dual-phase DUNE far detector will use a 12~m long extender of a similar design. Taken from \cite{ProtoDUNE-DP_Presentation}.}
\label{fig:ProtoDUNE-DP_HVExtender}
\end{figure}

The ProtoDUNE-DP cathode is biased at -300~kV \cite{ProtoDUNE-DP_Presentation} - twice the voltage used by (Proto)DUNE-SP. Although commercial HV power supplies that provide this bias are available, transmitting it all the way from the HV port at the top of the ProtoDUNE-DP cryostat to the cathode at the bottom - a distance of more than 8~m - is challenging. Rather than manufacturing a single, monolithic HV feedthrough for this purpose, a mechanically simpler solution is employed: a short feedthrough that uses the same design as that of WA105 (which, as noted previously, has been proven to reliably operate at a bias of -300~kV) coupled to a ``HV extender'' that effectively lengthens the feedthrough by 6~m \cite{ProtoDUNE-DP_Presentation}. The extender is a simple construction consisting of a conducting core surrounded by an insulator, with ``voltage degrader'' rings installed around the outside. These rings are connected to the outside of the field cage elements, and ensure that the electric field in the region outside the active TPC volume remains at zero. The installed extender is shown in Figure~\ref{fig:ProtoDUNE-DP_HVExtender}. DUNE-DP is envisioned to use an equivalent HV extender to connect its feedthrough and cathode, albeit one that is 12~m long. As an additional concern, the DUNE-DP cathode is required to operate at a bias of -600~kV, and this has necessitated the development (in collaboration with Heinzinger - the DUNE program's industrial partner in the production and procurement of HV power supplies) of a bespoke external power supply capable of producing such a high voltage, since no such device currently exists commercially \cite{DUNE_IDRv3}.

The maximum drift time in ProtoDUNE-DP will be 3.7~$\mu$s, and a goal of 7~ms electron lifetime has been set \cite{ProtoDUNE-DP_Presentation} (corresponding to 40~ppb $O_{2}$ equivalent impurities). As noted previously in Section~\ref{subsec:ProtoDUNE_SP}, the general requirements and designs for LAr provision, purification and circulation at ProtoDUNE-DP have been developed in collaboration with the larger single-phase experiments, and some of the systems are shared with ProtoDUNE-SP located nearby. Given the 20~ms lifetimes consistently achieved by ProtoDUNE-SP during its operation, it can be expected that ProtoDUNE-DP can and will reach, and likely exceed, its anticipated purity level.

\begin{figure}[h]
\vspace{2mm}
\centering
\includegraphics[width=0.60\textwidth]{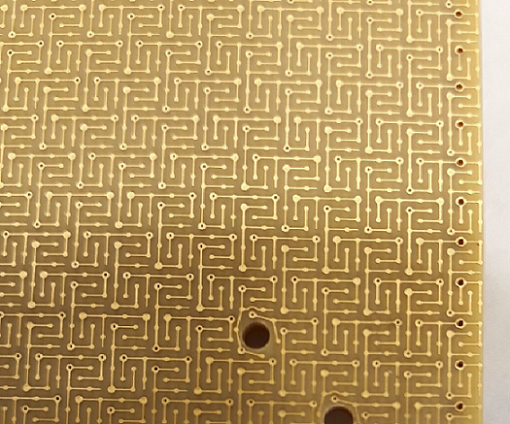}
\caption{Fine detail of the ProtoDUNE-DP anode plane surface. This design is optimised to give unambiguous 2D position reconstruction and equal charge sharing between the $x$ and $y$ views. An identical design has been used by WA105, and will also be used for the dual-phase DUNE far detector anode planes. Taken from \cite{ProtoDUNE-DP_Presentation}.}
\label{fig:ProtoDUNE-DP_Anode}
\end{figure}

Like its cathode, the 6 $\times$ 6~m ProtoDUNE-DP CRP is divided into 4 3 $\times$ 3~m units \cite{ProtoDUNE-DP_Presentation}. Each of these units is constructed using an identical design to the WA105 CRP described previously: an extraction grid of wires orientated in the $x$ and $y$ directions and covering the entire 3 $\times$ 3~m area of each unit, a LEM positioned 10~mm above the extraction grid and consisting of 50 $\times$ 50~cm sub-units (36 per CRP unit), and the anode plane - separated by 2~mm from the LEM and also comprised of 36 50 $\times$ 50~cm sub-units. The LEM units use the same physical characteristics as on WA105: 1~mm thickness, perforated by 500$\mu$m diameter holes with 800$\mu$m pitch, each with a 40$\mu$m dielectric rim. The surface design of the anode plane units, shown in Figure~\ref{fig:ProtoDUNE-DP_Anode}, is also identical to that of WA105. The same extraction and amplification fields (2 and 30~kV/cm respectively) are therefore also anticipated. In order to preserve the stability and flatness of each CRP unit across its entire surface, the LEM and anode plane units are mounted into a G10 fiberglass frame that spans the entire 3 $\times$ 3~m area, which is itself then suspended from an invar (iron-nickel alloy) frame \cite{ProtoDUNE-DP_Presentation}. This material is notable for its very low coefficient of thermal expansion \cite{DUNE_IDRv3} - meaning that it will undergo only a small contraction under cryogenic temperatures, and therefore that the overall deformation of the CRP as a whole will be minimised. The G10 frame is not directly attached to the invar frame in order to allow it to slide freely under its own thermal contraction without warping. The same CRP designs and components are anticipated to be used by DUNE-DP, but at a much larger scale: the 60 $\times$ 12~m CRP will require 80 of the 3 $\times$ 3~m units, and therefore 2880 50 $\times$ 50~cm LEM and anode plane units \cite{DUNE_IDRv3}.

\begin{figure}[ht]
\centering
\includegraphics[width=0.55\textwidth]{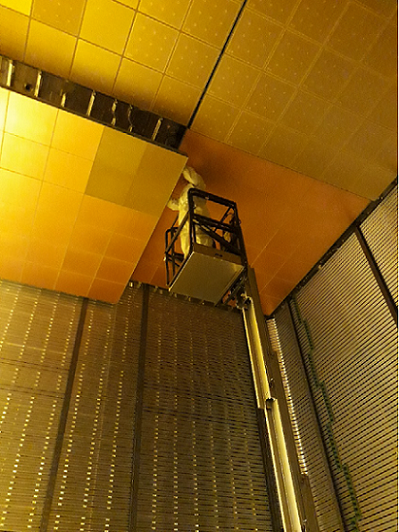}
\caption{Installation of the last of the four CRP units into the ProtoDUNE-DP TPC. Connection of the readouts to the corresponding signal feedthroughs was performed in-situ as each CRP unit was moved into its final position. Taken from \cite{ProtoDUNE-DP_Presentation}.}
\label{fig:ProtoDUNE-DP_CRPInstallation}
\end{figure}

Figure~\ref{fig:ProtoDUNE-DP_CRPInstallation} shows the installation of the final ProtoDUNE-DP CRP unit onto the TPC. As in WA105, each CRP unit's relative flatness and height is set via three adjustable cables passing into the TPC through dedicated feedthroughs. Readout is also performed using an identical electronics system to that of WA105: the anodes are connected to PCBs sealing the lower ends of the signal feedthroughs, the front-end electronics are housed within the vacuum-tight feedthrough chimneys, and the amplified and multiplexed signals are passed to warm electronics and the final acquisition and data storage systems located outside the cryostat \cite{ProtoDUNE-DP_Presentation}.

\begin{figure}[h!]
\vspace{5mm}
\centering
\includegraphics[width=0.90\textwidth]{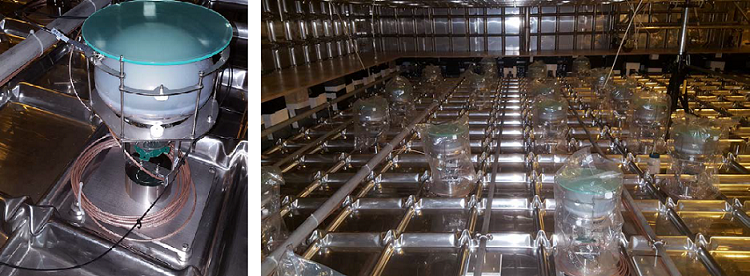}
\caption{(Left) One of the 36 8-inch diameter Hamamatsu R5912-MOD PMTs making up the ProtoDUNE-DP PDS, mounted in its own individual support structure that is glued directly to the interior surface of the detector cryostat. (Right) A view of the bottom of the cryostat, showing several of the PMTs installed. In both images, temporary plates are mounted above the PMTs for protection during installation. Both taken from \cite{ProtoDUNE-DP_Presentation}.}
\label{fig:ProtoDUNE-DP_PDS}
\end{figure}

Following from their proven operation on WA105, the ProtoDUNE-DP PDS consists of 36 8-inch diameter Hamamatsu R5912-MOD PMTs directly coated with TPB \cite{ProtoDUNE-DP_PDS}. These are located below the cathode, mounted within individual support structures that are glued directly to the interior surface of the cryostat - as shown in Figure~\ref{fig:ProtoDUNE-DP_PDS}. A stainless steel grid is positioned above the PMTs to act as a grounding plane, ensuring that there are no induced signals on the PMTs due to electrical cross-talk from the cathode. Detailed baseline and performance characterisations have been made on the PMTs prior to installation in the detector \cite{ProtoDUNE-DP_PDS}. As with other detector systems, this PDS scheme will be scaled up for use on DUNE-DP, with 720 PMTs expected to be used (assuming that DUNE-DP implements the same PMT ``density'' of 1 per m$^{2}$ of active area as in ProtoDUNE-DP) \cite{DUNE_IDRv3}.

Construction of ProtoDUNE-DP was completed in mid-2019, and this was followed by filling of the detector with LAr \cite{ProtoDUNE-DP_Presentation}. Initial testing and commissioning of all detector systems using cosmic muons was then performed beginning in late 2019, and has now been completed.

\subsection{Optical Readout and the ARIADNE Program}
Optical readout of dual-phase LArTPCs offers an alternative approach to the existing and previously described charge readout methodology. As noted previously in Section~\ref{sec:dualPhase}, Townsend multiplication occurring in the THGEM/LEM holes not only produces additional electrons, but also secondary scintillation light (hereafter denoted by ``S2''), and studies have shown that - depending on the exact amplification field across the THGEM/LEM - many hundreds of S2 photons can be produced for every drift electron \cite{Lightfoot2009, Monteiro2012, Buzulutskov2012, Buzulutskov2020, Aalseth2021}. Optical readout therefore relates to the detection of these S2 photons. 

Although there are a wide variety of photo-sensitive detectors available with which to observe the S2 light, one of the key requirements for use in a neutrino detector must be the capability for high resolution tracking and position reconstruction, as previously noted in Section~\ref{sec:dualPhase}. With this in mind, in recent years devices such as commercial SiPMs \cite{Lightfoot2009, Aalseth2021} and CCDs (``charge-coupled devices'') \cite{Mavrokoridis2014} have been used to optically read out small-scale dual-phase LArTPCs, and linearly graded SiPMs (``LG-SiPMs'') have also shown potential following initial tests conducted in a gaseous TPC \cite{ARIADNE_SiPM}. (LG-SiPMs are a relatively new type of SiPM that natively produces 3D position information using a combination of 2D microcell sensor structure and nanosecond timing resolution \cite{Gola2013}.) Much of the development of optical readout for dual-phase LArTPCs in the neutrino sector - including the aforementioned CCD and LG-SiPM experiments, and more recently the use of EMCCDs (``electron-multiplying CCDs'') and TPX3 (``Timepix3'') technology - has been performed by the ARIADNE (``\textbf{AR}gon \textbf{I}m\textbf{A}ging \textbf{D}etectio\textbf{N} chamb\textbf{E}r'') R\&D program \cite{ARIADNE_Website}, hosted at the University of Liverpool.

\begin{figure}[t!]
\centering
\includegraphics[width=0.75\textwidth]{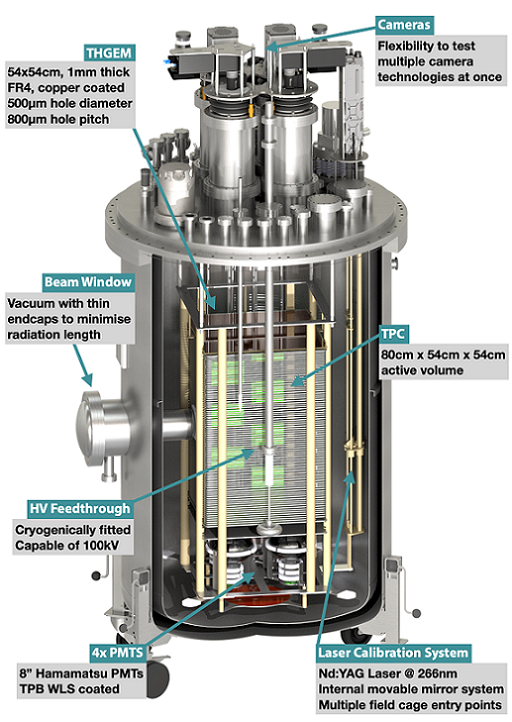}
\caption{A schematic depiction of the ARIADNE detector, with major components labelled. The TPB-coated reflector panels have been omitted for clarity. Taken from \cite{ARIADNE_Website}.}
\label{fig:ARIADNE_Detector}
\end{figure}

\begin{figure}[ht]
\centering
\begin{subfigure}[c]{0.45\textwidth}
\includegraphics[width=\textwidth]{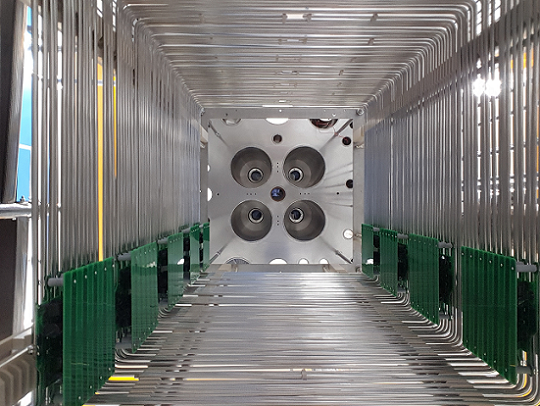}
\end{subfigure}
\hspace{1mm}
\begin{subfigure}[c]{0.45\textwidth}
\includegraphics[width=\textwidth]{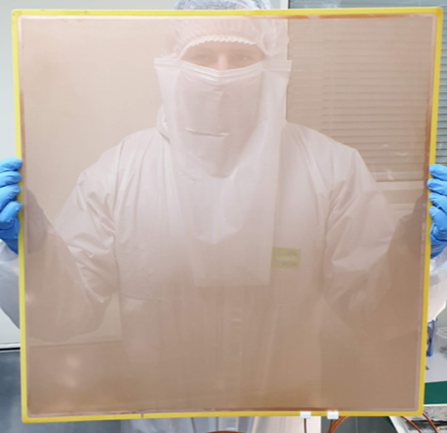}
\end{subfigure}
\caption{(Left) A view of the interior of the ARIADNE field cage, looking upwards from the position of the cathode. The field cage is defined by 79 stainless steel field-shaping rings, and the drift field is provided via the resistor chain, components of which are housed on the nine PCBs attached to the rings. Taken from \cite{ARIADNE_Website}. (Right) The THGEM used by ARIADNE, consisting of a a 1~mm thick FR4 panel with copper top and bottom surfaces, penetrated by 500~$\mu$m diameter holes with 800~$\mu$m pitch and 50~$\mu$m dielectric rims. The total hole coverage is 53 $\times$ 53~cm. Taken from \cite{ARIADNE_TDR}.}
\label{fig:ARIADNE_Components}
\end{figure}

The centrepiece of the ARIADNE program is the dual-phase ARIADNE LArTPC itself \cite{ARIADNE_TDR}, depicted in Figure~\ref{fig:ARIADNE_Detector}. At the centre of the detector is a 54 $\times$ 54 $\times$ 80~cm field cage, enclosing an active LAr mass of 330~kg and defined by 79 stainless steel field-shaping rings, with 1~cm centre-to-centre spacing. The rings are electrically connected via attached resistor chain elements. A view of the field cage interior is shown in Figure~\ref{fig:ARIADNE_Components} (left). The cathode, consisting of a tensioned stainless steel mesh held within a stainless steel frame, is positioned below the field-shaping rings and supplied with bias via a 2~m long HV feedthrough. The feedthrough design is identical to that used by other LArTPCs previously described, and contact between it and the cathode occurs via the feedthrough's spring-loaded tip touching the inside of a stainless steel torus welded to the cathode frame \cite{ARIADNE_TDR}.

The cathode grid has an optical transparency of 70\%, allowing the placement below it of four TPB-coated 8-inch diameter Hamamatsu R5912-MOD PMTs. As with other LArTPCs, the primary purpose of these PMTs is to detect prompt LAr scintillation light, the relative time of which denotes the event start and therefore allows the absolute $z$ position to be determined. In ARIADNE, the PMTs also provide supplementary S2 light detection capability alongside the primary optical readout devices. The overall light collection efficiency of the PMTs is enhanced by the placement of TPB-coated reflector panels around the outside of the field cage \cite{ARIADNE_TDR}.

The extraction grid is identical in design to the cathode, and located above the top field-shaping ring, 5~mm below the surface of the liquid. Bias to the extraction grid is provided via a dedicated HV feedthrough. The THGEM - shown in Figure~\ref{fig:ARIADNE_Components} (right) - is positioned 1~cm above the extraction grid, within the gas phase, and is constructed from a 1~mm thick 54 $\times$ 54~cm FR4 panel with copper top and bottom surfaces, penetrated by 500~$\mu$m diameter holes with 800~$\mu$m pitch and 50~$\mu$m dielectric rims, covering an area of 53 $\times$ 53~cm \cite{ARIADNE_TDR}. As many photo-sensitive detectors have high quantum efficiency in the visible wavelength range (and generally much lower efficiency at VUV wavelengths), a TPB-coated glass pane is mounted above the THGEM in order to shift the S2 light to wavelengths that are more conducive to detection.

\begin{figure}[h!]
\vspace{5mm}
\centering
\includegraphics[width=0.45\textwidth]{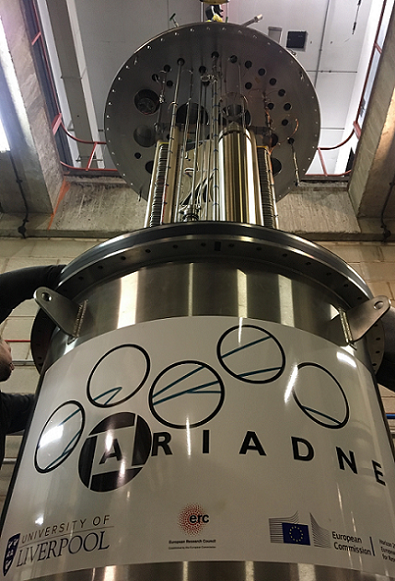}
\caption{Final installation of the fully assembled ARIADNE TPC into the cryostat, performed at the University of Liverpool. Taken from \cite{ARIADNE_Website}.}
\label{fig:ARIADNE_Construction}
\end{figure}

The internal detector components described thus far are contained within the ARIADNE cryostat - a double-walled construction utilising a combination of Mylar super-insulation and a vacuum jacket to maintain the interior cryogenic environment \cite{ARIADNE_TDR}. A vacuum-containing beam window penetrates both walls of the cryostat at the halfway height of the field cage, to allow incident beam particles to reach the active LAr with minimal material interaction. The top flange of the cryostat is a separate entity with its own vacuum jacket, and houses penetrations for the various feedthroughs and ports required for detector operation. Of particular note are the four DN200 CF chimneys and associated optical viewports, arranged in a 2 $\times$ 2 array around the flange's central axis, and onto which are mounted the primary optical readout devices (as shown in Figure~\ref{fig:ARIADNE_Detector}). When coupled with the appropriate optics, this position gives the devices an unobstructed view of the entire active THGEM area \cite{ARIADNE_TDR}, and the external location allows for easy access and installation of the readout devices. A view of the final installation of the fully assembled TPC into the cryostat is shown in Figure~\ref{fig:ARIADNE_Construction}.

Uniquely among the LArTPC detectors described in this paper, the main purification system of ARIADNE is located entirely within the cryostat, rather than externally \cite{ARIADNE_TDR}. The system consists of a filtration cartridge containing molecular sieves and copper meshes that serve to remove O$_{2}$ and H$_{2}$O impurities, connected to a custom-designed bellows pump that circulates the LAr between the cartridge and the surrounding internal cryostat volume.

\begin{figure}[h]
\vspace{2mm}
\centering
\includegraphics[height=0.30\textheight]{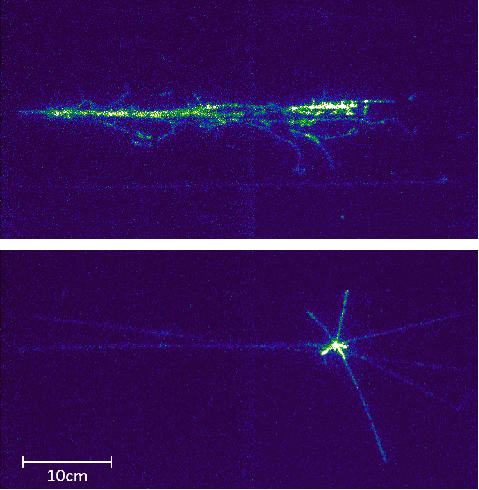}
\hspace{1mm}
\includegraphics[height=0.30\textheight]{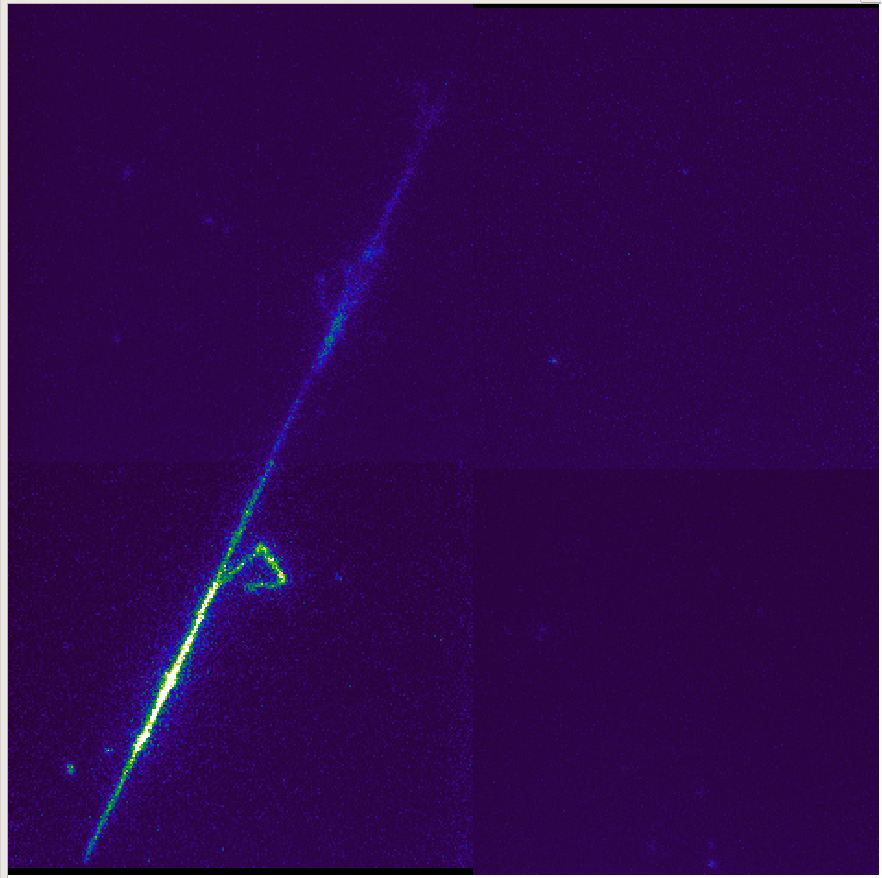}
\caption{Examples of events imaged by the ARIADNE detector using EMCCDs, showing (top left) a 5~GeV/c EM shower from a beam interaction, (bottom left) a 3~GeV/c antiproton candidate from a beam interaction, and (right) a cosmic muon interaction. All taken from \cite{ARIADNE_TDR}.}
\label{fig:ARIADNE_EMCCDEvents}
\end{figure}

The ARIADNE detector was first operated in 2018, using four Andor iXon 888 EMCCD cameras as the primary optical readout devices \cite{ARIADNE_TDR}. This followed from successful small-scale demonstration of the technology \cite{ARIADNE_EMCCDs}, which constituted the world first optical imaging of cosmic muons interacting in a dual-phase LArTPC. The EMCCDs have $\approx$ 80\% quantum efficiency at 430~nm, and are single-photon sensitive. They also possess built-in readout electronics, thereby significantly reducing the complexity of the detector readout systems, and can be operated with a variety of sensor size and pixel binning options - the choice of which affect both the per-pixel resolution and the maximum readout rate. During their operation on ARIADNE, a position resolution of 1.1~mm per pixel and a readout rate of 60~Hz were used \cite{ARIADNE_TDR}. The ARIADNE detector was initially placed on the CERN T9 beamline, where it was exposed to a variety of particle species covering a wide range of momenta, and this was followed by cosmic particle data-taking at the University of Liverpool. Figure~\ref{fig:ARIADNE_EMCCDEvents} shows examples of beam and cosmic interactions observed by ARIADNE using the EMCCDs, with the former representing the world first optical imaging of such events in LAr.

Though successful in proving the feasibility of the optical readout concept, operation of the ARIADNE detector using EMCCDs has demonstrated certain limitations with this technology. The most pressing are the necessity for reliable and unambiguous correlation with PMT signals in order to achieve full 3D position reconstruction (since the EMCCDs can only provide a 2D projection view of any given event), and the relatively low readout rate of the EMCCDs themselves \cite{ARIADNE_TDR}.

Beginning in 2019 and currently ongoing, the focus of the ARIADNE program has moved towards TPX3 based optical readout. This technology is a derivative of Medipix \cite{Medipix} - a suite of photo-sensitive pixel sensors originally developed to demonstrate how technologies developed for the LHC particle detectors could find application outside High Energy Physics. The TPX3 chip \cite{Timepix3_part1, Timepix3_part2, Timepix3_part3} itself consists of a photo-sensitive silicon sensor bump-bonded to an ASIC. The sensor is divided into a 256 $\times$ 256 array of 55~$\mu$m wide pixels, with each pixel being connected to its own dedicated circuitry in the ASIC. Incident photons are converted to electron/hole pairs, and the electrons are accumulated by each pixel's capacitor in the ASIC. The time taken for this accumulation to exceed a preset pixel-specific threshold is recorded (at 1.6~ns resolution) from an internal clock in the ASIC. This time is known as the ``time of arrival'' (ToA), and once the threshold is exceeded, that pixel is said to have been ``hit''. The capacitor is then discharged, and the additional time required after the ToA for the accumulation to fall back below threshold is called the ``time over threshold'' (ToT). This is proportional to the amount of accumulated charge, and therefore the number of photons incident on the pixel. Each hit pixel therefore provides four pieces of information upon readout: its $x$ and $y$ coordinates on the sensor surface, the ToA and ToT. In the context of imaging a three-dimensional particle track, the track's shape along the $z$ axis is related to the hit pixels' relative ToAs, and the summation of their ToTs is directly related to the total amount of LAr ionisation, and therefore the originating particle's energy. TPX3 technology therefore offers complete 3D position reconstruction and calorimetry from a single, self-contained device. Pixels that do not accumulate enough charge to exceed their thresholds are not read out - i.e. the data is naturally zero-suppressed, allowing for extremely fast readout rates of $\approx$ 80~MHits/s \cite{ARIADNE_TPX3LAr}.

As with any semiconductor device, the TPX3 chip is susceptible to thermal noise - typically $\approx$ 500 e$^{-}$ equivalent across the entire sensor surface \cite{Timepix3_part3}. However, in certain circumstances (such as in the presence of low energy particles or a low THGEM/LEM amplification field) there may not be enough incident S2 photons, and therefore electron/hole pairs, to overcome this minimum value. The S2 light must therefore first be amplified, and this is achieved using an image intensifier. Such devices generally consist of a photocathode (which converts the incident light into electrons), followed by one or more stages of electron multiplication, and a final phosphor layer that converts the electrons back into light. The commercial Photonis Cricket image intensifier \cite{PhotonisCricket} used by the ARIADNE experiment has a photocathode quantum efficiency of $\approx$ 30\% at 420~nm, a peak phosphor emission at 420~nm (matching well with the TPX3 chip's 90\% quantum efficiency at this wavelength), and an overall photon gain of 1.1 $\times$ 10$^{6}$ output photons per incident photon \cite{ARIADNE_TPX3LAr}.

\begin{figure}[h]
\centering
\includegraphics[width=0.55\textwidth]{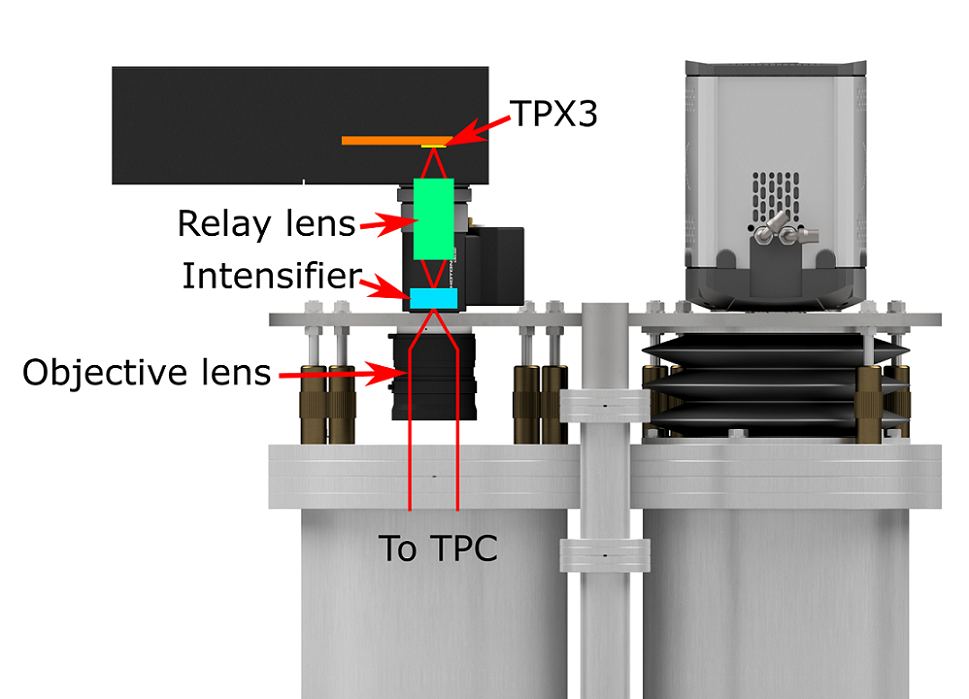}
\caption{Schematic depiction of the TPX3-based camera installed on the ARIADNE detector. An objective lens focuses the S2 light from the THGEM onto the photocathode face of the Photonis Cricket image intensifier. A built-in relay lens focuses the light produced by the intensifier's phosphor layer onto the TPX3 chip. The TPX3 housing (indicated by the black horizontal box) also contains back-end readout and acquisition electronics. The light-tight bellow around the objective lens has been omitted for clarity. Taken from \cite{ARIADNE_TPX3LAr}.}
\label{fig:ARIADNE_TPX}
\end{figure}

\begin{figure}[ht]
\centering
\includegraphics[width=0.48\textwidth]{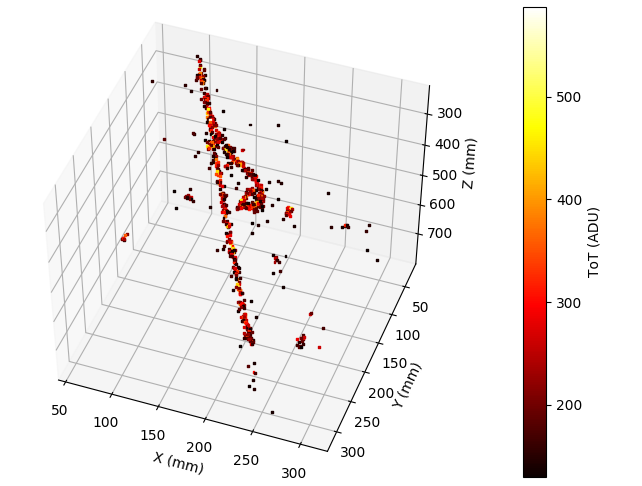}
\includegraphics[width=0.48\textwidth]{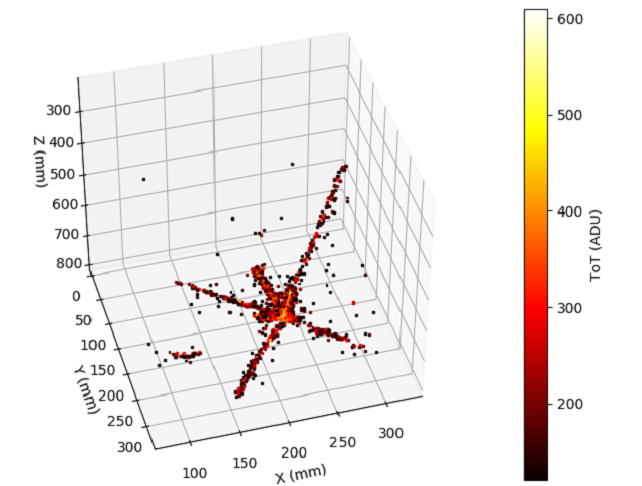}
\par\bigskip
\includegraphics[width=0.48\textwidth]{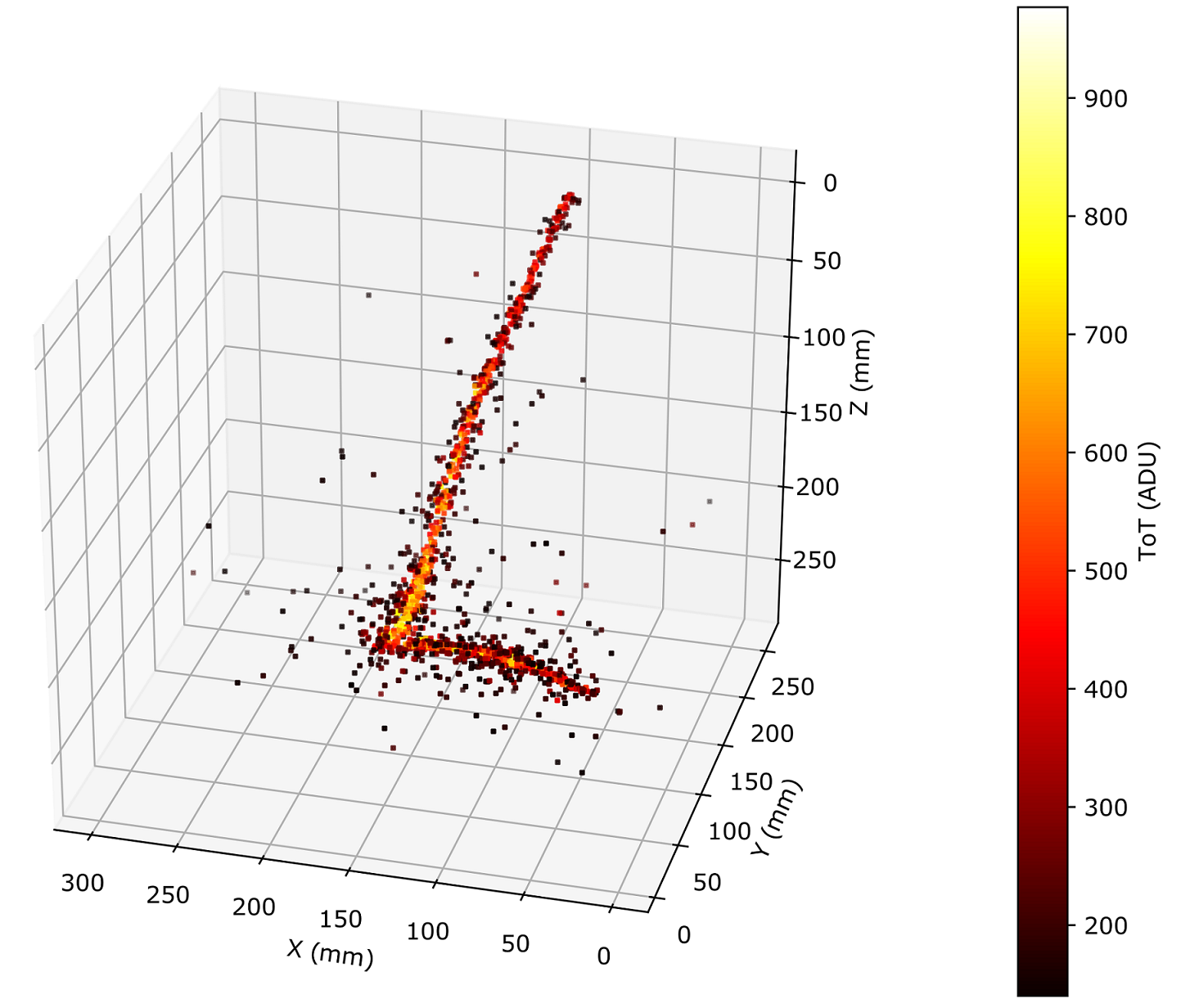}
\caption{Examples of cosmic particle interactions in LAr imaged by the ARIADNE detector using a TPX3-based camera, showing (top left) a through-going muon with delta ray, (top right) an antiproton candidate, and (bottom) a stopping muon and associated Michel electron. The colour scale represents the ToT per hit pixel. All taken from \cite{ARIADNE_TPX3LAr}.}
\label{fig:ARIADNE_TPXCosmics}
\end{figure}

Following from an initial proof-of-concept demonstration conducted on a smaller gaseous TPC \cite{ARIADNE_TPX3CF4}, a single TPX3-based camera - depicted in Figure~\ref{fig:ARIADNE_TPX} - has been installed on the ARIADNE detector, directly replacing one of the EMCCDs. Cosmic particle interactions have been imaged, and examples are shown in Figure~\ref{fig:ARIADNE_TPXCosmics}. The previously noted native zero-suppression of the data allows particle tracks to be clearly identified and reconstructed in 3D, regardless of their geometry and with no additional noise removal being required.

Based on this successful demonstration of an optical readout system utilising a single TPX3-based camera, a proposal has been made \cite{ARIADNE_LoIColdbox} for a larger-scale test at the CERN Neutrino Platform that will employ four TPX3-based cameras observing a larger, 2 $\times$ 2~m THGEM array. This will aim to further characterise the designs of the optics and electronics, and demonstrate the scalability of the optical readout concept.

\section{Summary and Outlook}
Single- and dual-phase LArTPCs have been, and continue to be, essential tools for the study of neutrinos. Central to all such detectors is the drifting of electrons produced through LAr ionisation due to the passage of charged particles, and the detection of these electrons (either directly or indirectly) in a way that allows both their population and 3D position to be unambiguously determined with high spatial and temporal precision.

In the field of single-phase detectors, the most well-established technology for achieving this is wire-based charge readout, generally taking the form of an anode plane assembly consisting of parallel planes of wires orientated so as to produce independent 2D views of particle interactions. These views are then combined to produce full 3D position reconstruction and calorimetric measurements. Many currently and soon-to-be operating single-phase LArTPCs - such as the ICARUS T600, MicroBooNE and SBND experiments of Fermilab's SBN Program and the ProtoDUNE-SP detector at CERN - make use of wire-based charge readout, alongside other proven technologies such as PMTs for observing the prompt LAr scintillation light and cold front-end electronics that can be located within the cryogenic environment, thereby benefiting from reduced noise and readout complexity. New approaches to single-phase LArTPC technology, and even the underlying design of such detectors themselves, are also being actively explored and promoted by a number of groups. These include the use of (X-)ARAPUCA light traps alongside or as an alternative to PMTs, pixelated charge readouts and modular TPCs as pioneered by the ArgonCube collaboration, and Vertical Drift's perforated PCB anodes and vertically orientated drift volume concept.

Dual-phase LArTPCs have only recently been given attention in the neutrino sector. However, due to having been used in dark matter searches for a number of years, the general detection principle of the dual-phase concept - multiplication of the drifted electrons using a THGEM/LEM or other micropattern detector located in the gaseous phase, before collection of the electrons on an anode plane - is well proven. The WA105 experiment was central to the initial establishment of dual-phase LArTPCs for neutrino detection, and its larger successor ProtoDUNE-DP aims to build on this work. In addition, the ARIADNE research program is currently investigating an alternative approach to dual-phase LArTPC design - that of optical readout, via the detection of the secondary scintillation light (initially using EMCCDs, and more recently with TPX3-based cameras) that is produced during electron multiplication. 

The culmination of many decades of LArTPC development and refinement is the DUNE program, which will have at its heart four colossal LArTPCs containing approximately 10,000~tons of active LAr each. At the time of writing, one of these detectors has been confirmed to use the established single-phase wire-based anode plane assembly design, with the remaining three still to be decided upon. However, given the well-proven nature of both the single- and dual-phase concepts, as well as the variety and potential of the many new approaches being explored, it is clear that no matter what technologies are eventually used, LArTPCs will continue to play a critical and indispensable role in the neutrino sector for many years to come.

\vspace{2mm}
\noindent \hrulefill
\vspace{2mm}

\authorcontributions{Conceptualization, K.Mj. and K.Mv.; Investigation, K.Mj. and K.Mv.; Writing, K.Mj. and K.Mv. All authors have read and agreed to the published version of the manuscript.}

\funding{The authors are supported by European Research Council Grant No. 677927.}

\acknowledgments{The authors would like to thank the reviewers for the time and effort they have taken to respond to this manuscript, as well as their specific and direct feedback. We found the comments and suggestions most helpful.}

\conflictsofinterest{The authors declare no conflict of interest. The funders had no role in the choice of contents, writing, or decision to publish the manuscript.}

\reftitle{References}

\end{document}